\documentclass{resources/aa}
\usepackage{graphicx}
\usepackage[varg]{txfonts}
\usepackage{physics}
\usepackage{amsmath, amssymb}
\usepackage{dsfont}
\usepackage{bbm}
\usepackage{xcolor}
\usepackage{ulem}
\usepackage{xpatch}
\usepackage{hyperref}
\usepackage{import}
\usepackage{makeidx}
\usepackage{changes}
\usepackage{hhline}
\usepackage{natbib}
\usepackage{placeins}
\usepackage{stfloats}
\defcitealias{Cecil2024b}{Paper~I}
\defcitealias{Cecil2026}{Paper~II}

\AtBeginDocument{%
  \setlength{\abovedisplayskip}{6pt}%
  \setlength{\belowdisplayskip}{6pt}%
  \setlength{\abovedisplayshortskip}{6pt}%
  \setlength{\belowdisplayshortskip}{6pt}%
}

\usepackage{array}
\usepackage{makecell}
\usepackage{hhline}

\newcommand{\colout}[1]{\bgroup\markoverwith{\textcolor{#1}{\rule[.5ex]{2pt}{0.4pt}}}\ULon}

\makeatletter
\renewcommand*\aa@pageof{, page \thepage{} of \pageref*{LastPage}}
\newcommand{\pder}[2][]{\frac{\partial#1}{\partial#2}}
\makeatother

\renewcommand{\arraystretch}{1.5}

\makeindex

\begin{document}

\titlerunning{Inner dead zone instability tied to magnetic fields}
\authorrunning{M. Cecil et al.}
\title{MRI-triggered instability at the inner dead zone edge: \\ disc evolution and burst modes tied to magnetic field strengths } 

\author{
 M.~Cecil\inst{1 \and 2}\thanks{Corresponding author; \texttt{cecil@mpia.de}},
 M.~Flock\inst{1}, D.~Steiner\inst{3}
}

\institute{
 Max Planck Institute for Astronomy (MPIA), Königstuhl 17, 69117 Heidelberg, Germany
 \and
 Fakultät für Physik und Astronomie, Universität Heidelberg, Im Neuenheimer Feld 226, 69120 Heidelberg, Germany \and
 Department of Astrophysics, University of Vienna,
 Türkenschanzstrasse 17, A-1180 Vienna, Austria
}

\date{Received ....; accepted ....}

\abstract
{The inner edge of the dead zone (DZIE) in protoplanetary discs is prone to periodic instability caused by the activation of the magneto-rotational instability (MRI) within the weakly turbulent regions. Capturing the triggering and evolution of the instability mechanism, along with the resulting accretion burst signatures, requires coupling MRI activation to the local magnetic field via non-ideal magnetohydrodynamic (MHD) effects.}
{Our study shows how different large-scale magnetic field configurations set the structure of the inner disc and regulate the resulting evolution and morphologies of periodic instability cycles. }
{We performed 2D and 3D radiation hydrodynamic simulations of the regions around the DZIE of a Class II disc over timescales of $10^3$\;yr. We significantly extended previous studies by implementing MRI activation criteria based on ambipolar and Ohmic diffusion and by prescribing magnetic field strength profiles comprising stellar and disc components. }
{The frequency, shapes and consequences of the episodic accretion events are highly sensitive to the magnetic field strength in the inner disc. We recover previously reported dynamic behaviour by considering relatively strongly magnetised discs. A new burst mode is revealed, in which the MRI active region cannot expand far into the disc in the presence of weak magnetic fields. In this narrow burst mode, the pressure maximum at the inner dead zone edge does not remain static even during quiescence. A distinct dichotomy between the wide and narrow burst modes is established by the hydrodynamic (in)stability of the ionisation front. Both modes are additionally separated into a reflaring and a non-reflaring version, mostly determined by the stellar magnetic field strength. Our setup does not lead to the emergence of classical thermal instability by hydrogen ionisation. The structure of the MRI active region in quiescence changes from a simple radial MRI transition to a layered structure that converges towards the midplane near the star. Our 3D model reveals the breaking of the density features produced in the narrow burst mode, leading to strong vortices at radii smaller than 0.5~AU.}
{Coupling MRI activity directly to different magnetic field strengths via non-ideal MHD effects, rather than using simple temperature thresholds, enables a variety of burst modes. Each mode exhibits characteristic accretion burst signatures and produces vastly different consequences for the evolution of the inner disc structure and the conditions of planet formation and migration.}


\maketitle

\section{Introduction} \label{sec:introduction}
The terrestrial planets of the Solar System and the majority of currently detected exoplanets are located in close proximity to their host star \citep[with observational selection effects, e.g.,][]{Vanderburg2016, Hara2023, Ananyeva2023}. To understand the conditions under which these planets formed at (or migrated to) these locations requires studying the structure and evolution of the inner regions of protoplanetary discs (PPDs). They feature a transition between a gaseous, ionised region closest to the star, where the magnetorotational instability \citep[MRI,][]{Balbus1991} can be sustained, and a weakly turbulent MRI-dead zone \citep[e.g.,][]{Dzyurkevich2013, Flock2016}. The dead zone's inner edge (DZIE) hosts a local pressure maximum, which can act as an efficient trap for solid particles and migrating planets \citep{Dzyurkevich2010, Flock2019, Chrenko2022}. However, the inner disc is prone to significant variability, which often translates into inconstancy of accretion onto the host star \citep[e.g.,][]{Lin1985, Zhu2009b, Zhu2010a, Audard2014, Kadam2019, Elbakyan2025}. In our previous work \citep[][hereafter \citetalias{Cecil2024b} and \citetalias{Cecil2026}]{Cecil2024b, Cecil2026}, we showed that the presence of the DZIE can intrinsically lead to periodic instability of the inner disc, manifesting as episodic accretion outbursts, by activating the MRI in the dead zone \citep[MRI outbursts, see also e.g.,][]{Zhu2009, Kadam2020, Cleaver2023}. The consequential restructuring of the inner disc leads to the periodic destruction of the pebble trap at the DZIE while new transient pressure maxima are formed throughout the innermost disc, in which dust can accumulate and grow \citep{Dullemond2018, Lee2022, Ziampras2026}. \par
The MRI-triggered instability mechanism analysed in \citetalias{Cecil2024b} and \citetalias{Cecil2026} relies on the trapping of heat near the midplane as material piles up at the DZIE due to differential angular momentum transport between the permanently MRI active region and the dead zone. The outburst is ignited as soon as the conditions for MRI activation in the dead zone are satisfied, leading to a runaway heating process. Within the adopted simulation setup, the temperatures never reach the regime of hydrogen ionisation, which would lead to the classical thermal instability often associated with outbursts in high-mass discs \citep{Bell1993, Nayakshin2024, Elbakyan2024, Elbakyan2025}. As discussed in \citetalias{Cecil2026}, this is partly due to a thermostat effect resulting from dust sublimation \citep[e.g.,][]{Woitke2024}. However, younger and more massive discs than those considered in our models could overcome this effect by sublimating the entire dust content near the midplane, which lets the temperatures increase further towards the regime of hydrogen ionisation, especially at small distances from the star (as indicated in the model \texttt{M3} discussed in \citetalias{Cecil2024b} and the S-curve analysis at small radii in \citetalias{Cecil2026}). For an investigation into the role of classical thermal instability, we refer to \citet{Nayakshin2024}, who consider more massive discs and distances down to the stellar surface within one-dimensional models.\par
We emphasise that our studies are not intended to reproduce specific accretion outburst signatures observed in, for instance, FU Ori- or EX Lup-type objects, which are summarised in compilations such as the outbursting young stellar objects catalogue \citep[\texttt{OYCAT}][]{Pea2025}. Instead, we aim to show the variable evolution of the inner regions around the DZIE of an early Class II-type disc and to analyse its consequences for the inner disc structure, planet formation conditions and stellar accretion rates.  \par
Our previous models indicate that two essential aspects control the MRI-triggered instability mechanism: (i) the opacity of the disc material regulating the heat transfer, and (ii) the conditions for MRI activation. In \citetalias{Cecil2026}, we showed that although different gas and dust opacity descriptions have a significant effect on the inner disc structure, they do not alter the evolution of the instability or the morphology of the resulting outbursts (apart from the quiescent timescales). In this work, we aim to determine whether elaborate descriptions of MRI activation can change this picture. \par
In \citetalias{Cecil2024b} and \citetalias{Cecil2026}, we implemented a simple activation temperature threshold, $T_\mathrm{MRI}$, as commonly used in previous dynamical studies \citep[e.g.,][]{Zhu2010a, Bae2013, Macfarlane2019, Steiner2021, Chambers2024}. However, the ionisation state in the inner disc and the efficiency of non-ideal magnetohydrodynamic (MHD) effects, which ultimately determine whether the MRI is active or not, are functions of not just the temperature \citep{Desch2015, Williams2025}. In particular, MRI activity is crucially affected by the presence and strength of magnetic fields in the disc, a dependence that is the focus of this work. \par
Observations of magnetic field strengths in PPDs have proven to be challenging. Recently, \citet{Teague2025} reported measurements of a radially resolved magnetic field structure, yielding strengths of around 10~mG at a distance of $\sim\!\!60$~AU from the central star. At smaller radii, constraints on magnetic field strengths can be derived for the protosolar nebula from paleomagnetic studies of meteorites, whose parent bodies formed in the inner regions \citep[e.g.,][]{Weiss2021, Fu2023}. The oldest samples from which the remnant magnetisation can be traced back to an inner disc magnetic field point to strengths of >0.4~G at distances of 1--7~AU from the protosun \citep{Fu2014, Fu2021, Maurel2024, Mansbach2024}.\par
Since the bursting mechanism produces strong density and pressure gradients, a variety of thermo-hydrodynamic instability processes might be expected, among which are Rayleigh instability \citep{Chandrasekhar1961, Yang2010}, Convective Overstability \citep{Klahr2014, Latter2016, Teed2021} and Rossby Wave Instability \citep[RWI][]{Lovelace1999, Lovelace2014}. To follow up on the stability analysis of these features conducted in \citetalias{Cecil2026}, non-axisymmetric simulations are required. \par
We present 2D and 3D radiation hydrodynamic simulations of the inner disc over a thousand-year timescale, including irradiation by the central star, radiative transport in the flux-limited diffusion approximation, dust sublimation, and detailed gas and dust opacity descriptions. We determine the local MRI activity based on ambipolar and Ohmic diffusion via pre-calculated tables of diffusivities provided by the thermo-chemical ionisation model of \citet{Desch2015}. To retain computational feasibility, we do not employ full MHD models and track magnetic flux transport \citep[as recently conducted in the inner disc by e.g.,][]{Iwasaki2024, Roberts2025, Roberts2026}, but prescribe profiles of magnetic field strengths, informed by recent steady-state solutions \citep{Dudorov2014, Steiner2025}. \par
The paper is organised as follows. Section \ref{sec:method} summarises the numerical and physical setup of our simulations. The main results are presented in Sec. \ref{sec:results} before we derive the most important implications in Sect. \ref{sec:discussion}. The conclusions are given in Sect. \ref{sec:conclusion}.

\section{Method}\label{sec:method}

The general structure of the simulations conducted in this work closely follows the methods described in \citetalias{Cecil2024b} and \citetalias{Cecil2026}. The most important governing equations adopted from our previous work are briefly summarised again in Sect.~\ref{sec:meth_eqs}. The central extension of the models presented in this work is the description of the viscosity, which is laid out in Sect. \ref{sec:meth_viscosity}. The imposed magnetic field is described in Sect. \ref{sec:meth_Bfield}. Sections \ref{sec:meth_diffusivities} and \ref{sec:meth_nonthermal} present the considerations for including non-ideal diffusion coefficients and non-thermal ionising sources, respectively. Finally, the numerical details of our models are given in Sect. \ref{sec:numerical}.
\subsection{Governing equations} \label{sec:meth_eqs}

Analogous to \citetalias{Cecil2024b} and \citetalias{Cecil2026}, the structure and evolution of the inner PPD in our models followed the set of coupled radiation hydrodynamic equations,
\begin{align}
    &\pder[\rho_\mathrm{g}]{t} \, + \nabla \cdot ( \rho_\mathrm{g} \, \vec v ) = 0\;, \label{eq:cont} \\
    &\pder[\rho_\mathrm{g} \, \vec v]{t} + \nabla \cdot (\rho_\mathrm{g} \, \vec v  \, \vec v^T) + \nabla P_\mathrm{g} = -\rho_\mathrm{g} \, \nabla \Phi + \nabla \cdot \vec \Pi \;, \label{eq:mot} 
\end{align}
\vspace{-0.4cm}
\begin{multline}
    \pder[E]{t} + \nabla \cdot [(E + P_\mathrm{g}) \, \vec v] = -\rho_\mathrm{g} \, \vec v \cdot \nabla \Phi - \vec \Pi : \nabla \vec v  \\ 
    - \kappa_\mathrm{P} \, \rho_\mathrm{g} \, c \, (a_\mathrm{R}  \, T_\mathrm{g}^4 - E_\mathrm{R}) - \nabla \cdot F_\mathrm{irr}  \;, \label{eq:ene} 
\end{multline}
\vspace{-0.4cm}
\begin{align}
    &\frac{1}{\gamma -1} \, \pder[P_\mathrm{g}]{t}=-\kappa_\mathrm{P} \, \rho_\mathrm{g} \, c \, (a_\mathrm{R} \, T_\mathrm{g}^4 - E_\mathrm{R}) - \nabla \cdot F_\mathrm{irr} \; , \label{eq:rad1} \\
    &\pder[E_\mathrm{R}]{t} - \nabla \cdot \left ( \frac{c \, \lambda}{\kappa_\mathrm{R} \, \rho_\mathrm{g}} \, \nabla E_\mathrm{R} \right ) = \kappa_\mathrm{P} \, \rho_\mathrm{g}\, c \, (a_\mathrm{R} \, T_\mathrm{g}^4 - E_\mathrm{R}) \;, \label{eq:rad2} 
\end{align}
\noindent where Eqs. \ref{eq:cont}--\ref{eq:ene} are the continuity, momentum and total energy equations, respectively, and Eq. \ref{eq:rad1} and \ref{eq:rad2} describe the radiative transport in the flux-limited diffusion approximation. $\rho_\mathrm{g}$ denotes the gas mass density, $\vec v=(v_r, v_\theta, v_\mathrm{\varphi})$ the velocity vector in spherical coordinates, $P_\mathrm{g}$ the gas pressure, $\Phi$ the gravitational potential, $\vec \Pi$ the viscous stress tensor, $E$ the total energy, $T_\mathrm{g}$ the gas temperature, $E_\mathrm{R}$ the radiation energy density and $F_\mathrm{irr}$ the irradiation flux. In addition, $\kappa_\mathrm{P}$ and $\kappa_\mathrm{R}$ represent the Planck and Rosseland mean opacities, $c$ and $a_\mathrm{R}$ are the speed of light and the radiation constant, respectively, $\gamma=1.42$ is the adiabatic factor and $\lambda$ is the flux--limiter function \citep{Levermore1981}. The ideal gas equation of state served as a closure relation. \\
Following \citetalias{Cecil2026}, we used the DIANA standard dust opacities, $\kappa_\mathrm{P,d}$ and $\kappa_\mathrm{R,d}$ \citep{Woitke2016}, and the gas opacities, $\kappa_\mathrm{P,g}$ and $\kappa_\mathrm{R,g}$ from \cite{Malygin2014}. The total Planck or Rosseland opacity was then constructed utilising the dust-to-gas mass ratio $f_\mathrm{D2G}$,
\begin{equation}
\kappa_\mathrm{P,R}=\kappa_\mathrm{P,R,d}+f_\mathrm{D2G}\kappa_\mathrm{P,R,g} \;.
\end{equation}
\noindent In the description of $f_\mathrm{D2G}$, we accounted for the sublimation of silicates analogously to \citet{Isella2005}. For details, we refer to Appendix A of \citetalias{Cecil2026}. Since the investigations in \citetalias{Cecil2026} did not show a notable influence of frequency-dependent irradiation on the occurrence and evolution of accretion burst cycles, we chose to describe the stellar irradiation in the grey approximation\footnote{In \citetalias{Cecil2026}, we varied the stellar temperature according to the variability of the stellar luminosity due to the added accretion shock luminosity. However, since this variability had no significant effect on the accretion burst mechanism, we kept $T_\star$ constant in the models of this work. }\par

The viscous stress tensor was given as,
\begin{equation}
    \vec \Pi=\mu \left [ \nabla \vec v + (\nabla \vec v)^\mathrm{T}-\frac{2}{3}(\nabla \cdot \vec v)\vec {\mathrm{I}} \right ] \; ,
\end{equation}
\noindent where $\vec {\mathrm{I}}$ is the unit tensor and,
\begin{equation}
\mu=\rho_\mathrm{g}\nu=\rho_\mathrm{g}\frac{\alpha\,c_\mathrm{s}^2}{\Omega}\;,
\end{equation}
is the dynamic viscosity with the kinematic viscosity $\nu$ following the standard $\alpha$--prescription \citep{Shakura1973}, the speed of sound $c_\mathrm{s}=\sqrt{\gamma P_\mathrm{g}/\rho_\mathrm{g}}$, the orbital frequency $\Omega$ and the stress-to-pressure ratio $\alpha$. The structure of $\alpha$ is at the core of this work's models and is described in Sect. \ref{sec:meth_viscosity}. \par
We constructed the initial models by solving the equations for vertical hydrostatic equilibrium together with Eqs. \ref{eq:rad1} and \ref{eq:rad2}. The surface density of the hydrostatic model was calculated as,
\begin{equation} \label{eq:Sigma}
    \Sigma=\frac{\dot{M}_\mathrm{init}}{3\, \pi \, \nu} ~,
\end{equation} 
\noindent with $\dot{M}_\mathrm{init}$ being a constant mass transport rate through the disc. 
Note that in contrast to our previous work, the radiative transport equations do not include the source term describing the viscous heat dissipation. As shown in \citetalias{Cecil2024b}, including viscous heating in the establishment of the hydrostatic models can lead to strong oscillations in the density and temperature structure due to the multiplicity of possible solutions. These solutions are part of the limit cycle characterising the outburst behaviour, meaning that the initial model already satisfies the conditions for entering the burst phase. To avoid these oscillations and to ensure that the various models in this work can be fairly compared by starting from the same initial surface density structure, we excluded the viscous heating term from the initial hydrostatic models.\par
By solving the set of Eqs. \ref{eq:cont}--\ref{eq:rad2}, we then followed the disc's evolution, starting from the hydrostatic structure. For all simulations, we used the PLUTO code \citep{Mignone2007} together with the flux-limited diffusion radiative transport description of \cite{Flock2013}.

\subsection{Viscosity} \label{sec:meth_viscosity}
In our previous work, we described the viscous $\alpha$ parameter by a smooth transition between the small dead zone value, $\alpha_\mathrm{DZ}$, and the larger value in the MRI active regions, $\alpha_\mathrm{MRI}$, around a certain MRI activation threshold temperature $T_\mathrm{MRI}$. In this study, we adopted a different approach to determine where the MRI is active and where it is suppressed. In the ideal case, full 3D MHD simulations \citep[e.g.,][]{Flock2017a} are required for this purpose. However, since we intend to investigate accretion burst cycles originating from the very inner disc and requiring simulation times of potentially over a thousand years for each model, this method is computationally prohibitive. Instead, we adopted detailed criteria for MRI activation that have been commonly used in previous studies \citep[e.g.,][]{Dzyurkevich2013, Mohanty2018, Jankovic2021, Delage2022}. \par 
The MRI activity was determined by accounting for the non-ideal MHD effects of ambipolar and Ohmic diffusion. The two criteria that need to be fulfilled in order for the MRI to be active and saturated can be written as,
\begin{align}
    &\Lambda>1 \;,\label{eq:MRIcrit1} \\
    &\beta>\beta_\mathrm{min}(\mathrm{Am)} \; , \label{eq:MRIcrit2}
\end{align}
\noindent where $\Lambda$ is the Ohmic Elsässer number,
\begin{equation}
    \Lambda=\frac{v_\mathrm{A,z}^2}{\eta_\mathrm{O}\Omega_\mathrm{K}} \; ,
\end{equation}
\noindent and $\beta_\mathrm{min}(\mathrm{Am})$ is the minimally necessary value of the plasma $\beta=8\pi P_\mathrm{g}/|\vec{\mathrm{B}}|^2$ parameter for the MRI to develop \citep{Bai2011},
\begin{equation}
    \beta_\mathrm{min}(\mathrm{Am})=\left [\left( \frac{50}{\mathrm{Am}^{1.2}} \right)^2+\left( \frac{8}{\mathrm{Am}^{0.3}}+1 \right)^2 \right]^{1/2}  \;.
\end{equation}
Equation \ref{eq:MRIcrit2} ensures that the magnetic field is not strong enough to quench the MRI activity by ambipolar diffusion, the importance of which is characterised by the ambipolar Elsässer number, 
\begin{equation}
    \mathrm{Am}=\frac{v_\mathrm{A}^2}{\eta_\mathrm{AD}\Omega_\mathrm{K}} \;.
\end{equation}
\noindent The Alfvén velocity is given by, 
\begin{equation}
    v_\mathrm{A}=\frac{|\vec{\mathrm{B}}|}{\sqrt{4\pi\rho_\mathrm{g}}} \;, ~~~~~v_\mathrm{A,z}=\frac{B_\mathrm{z}}{\sqrt{4\pi\rho_\mathrm{g}}} \;,
\end{equation}
with $v_\mathrm{A,z}$ being its vertical component. Since we only consider purely vertical magnetic fields in our models, we can set $v_\mathrm{A}=v_\mathrm{A,z}$. The description of the magnetic field $\vec{\mathrm{B}}$ is given in section \ref{sec:meth_Bfield}. The diffusivities of ambipolar and Ohmic diffusion are denoted by $\eta_\mathrm{AD}$ and $\eta_\mathrm{O}$, respectively, and are explained in Sect. \ref{sec:meth_diffusivities}. \par
If both of the criteria given in Eqs. \ref{eq:MRIcrit1} and \ref{eq:MRIcrit2} are fulfilled, the $\alpha$ parameter entering the viscous stress tensor adopts the MRI active value $\alpha_\mathrm{MRI}$. Otherwise, the MRI is suppressed and $\alpha$ is described by a dead zone value $\alpha_\mathrm{DZ}$. In order to stay consistent with our previous work, we chose $\alpha_\mathrm{MRI}=10^{-1}$ and $\alpha_\mathrm{DZ}=10^{-3}$.

\subsection{Magnetic field description} \label{sec:meth_Bfield}
The goal of this work is to investigate the inner disc instability mechanism and possible burst morphologies in response to different magnetic field strength profiles. The only purpose of the magnetic field description in our models is to help determine the regions of MRI activity. Other effects of the magnetic field, such as torques or winds, are not included in this study. Additionally, we aimed to preserve the comparability of the different models and to avoid further complications in the analysis arising from complex magnetic field structures and their evolution. Therefore, the magnetic field in our models should be regarded as a fossil field, without contributions from potential dynamo mechanisms in the disc. We do not follow the typical, optimistic approach of choosing the magnetic field strength to maximise MRI activity \citep[e.g.,][]{Jankovic2021, Delage2022}. Instead, we prescribe the large-scale magnetic field threading the disc as purely vertical (hence only dependent on the distance from the star) and consisting of two components,
\begin{equation}
    \vec{\mathrm{B}}=B_\mathrm{z}(r)=B_\mathrm{dipole}(r)+B_\mathrm{disc}(r) \; .
\end{equation}
\noindent The dipolar component $B_\mathrm{dipole}(r)$ represents the stellar magnetic field with a strength of $B_\star$ at the stellar surface,
\begin{equation}
    B_\mathrm{dipole}(r)=B_\star \left( \frac{R_\star}{r}\right)^3 \;,
\end{equation}
\noindent where $R_\star$ is the stellar radius. For $B_\star$, we adopt values typical for classical T-Tauri stars \citep[e.g.,][]{Johns2007}. \par
The recent study of \cite{Steiner2025} has shown that in a steady-state solution, the magnetic field in the inner disc is mostly dominated by the stellar dipole component up to a distance of about 0.5 AU, after which the vertical magnetic field strength levels off towards an approximately constant value at a radius of several AU in the dead zone. To approximate this behaviour, we chose a radially constant disc field component $B_\mathrm{disc}(r)=\mathrm{const.}$ for several models in this study. To also test the effect of a relatively strong, radially decreasing disc field \citep[e.g.,][]{Dudorov2014}, we conducted additional simulations including a disc component of $B_\mathrm{disc}(r)=0.1/\left(r\;\mathrm{[AU]}\right)$.

\subsection{Diffusivities} \label{sec:meth_diffusivities}
The MRI activation threshold temperature used in our previous work was informed by investigations of ionisation levels in high-temperature regions of discs conducted by \cite{Desch2015}. However, the ionisation state and, therefore, the diffusion coefficients, $\eta_\mathrm{AD}$ and $\eta_\mathrm{O}$, are influenced by more than just the temperature of the gas, especially considering that thermionic emission plays a major role in determining ionisation levels in addition to collisional ionisation of alkali metals. For the models in this work, we used lookup tables of $\eta_\mathrm{AD}$ and $\eta_\mathrm{O}$ as functions of gas density, temperature, plasma $\beta$, and non-thermal ionisation rate, based on the work of \cite{Desch2015}. The tables are valid for a dust-to-gas ratio of $10^{-3}$ (which matches the maximum value, $f_0$, in our models) and do not account for the dependence on the dust size distribution\footnote{For more recent models taking the varying conditions of the dust distribution into account, we refer to \cite{Williams2025}}. Using $\rho_\mathrm{g}$, $T_\mathrm{g}$ and $P_\mathrm{g}$ from our radiation hydrodynamic solution, the magnetic field strength from our prescription (Sect. \ref{sec:meth_Bfield}) and the non-thermal ionisation rate $\zeta$ based on the description given in Sect. \ref{sec:meth_nonthermal}, we interpolate the diffusion coefficients for every cell in the computational domain at every timestep. For input values outside the tables' covered range, we apply logarithmic extrapolation (assuming power-law relationships between the input values and the diffusion coefficients). The sensitivity of our results to this extrapolation is explored in Appendix \ref{sec:app_extrapolation}.

\subsection{Non-thermal ionisation rate} \label{sec:meth_nonthermal}
For the evaluation of the total non-thermal ionisation rate, we adopted methods commonly applied in previous works \citep[e.g. ][]{Dzyurkevich2013, Jankovic2021, Delage2022, Steiner2025}. Hereby, we considered ionisation contributions from radioactive decay, $\zeta_\mathrm{RD}$, incident cosmic rays, $\zeta_\mathrm{CR}$ and stellar X-rays, $\zeta_\mathrm{X}$. The total non-thermal ionisation rate was calculated as the sum of the individual contributions,
\begin{equation}
    \zeta=\zeta_\mathrm{RD}+\zeta_\mathrm{CR}+\zeta_\mathrm{X} \;.
\end{equation}
Considering the decay of radionuclides, we set $\zeta_\mathrm{RD}=7.6\,\cdot\,10^{-19}\,\mathrm{s}^{-1}$ \citep{Umebayashi2009}, neglecting the potential small variations with radius \citep{Cleeves2013} or dust-to-gas mass ratio \citep{Delage2022}. \par
$\zeta_\mathrm{CR}$ and $\zeta_\mathrm{X}$ are the cosmic ray and X-ray ionisation rate and were calculated as functions of the cylindrical vertical column density, $\Sigma(R,Z)=\int_Z^\infty\rho_\mathrm{g}(R,Z') \,\mathrm{d}Z'$ and $\Sigma(R,Z)=\int_{-\infty}^Z\rho_\mathrm{g}(R,Z') \,\mathrm{d}Z'$ for the upper and lower hemisphere, respectively, with $R$ and $Z$ representing the cylindrical radius and height. \par
The cosmic ray ionisation rate is given by \citep{Umebayashi2009},
\begin{equation}
    \zeta_\mathrm{CR}(R,Z)=\frac{\zeta_\mathrm{CR,ISM}}{2}e^{-\frac{\Sigma(R,Z)}{\lambda_\mathrm{CR}}} \left( 1+\left( \frac{\Sigma(R,Z)}{\lambda_\mathrm{CR}}\right)^\frac{3}{4}\right)^{-\frac{4}{3}} ~~~,
\end{equation}
\noindent where $\zeta_\mathrm{CR,ISM}=10^{-17}\,\mathrm{s}^{-1}$ is the unattenuated ISM cosmic ray ionisation rate and $\lambda_\mathrm{CR}=96\,\mathrm{g\,cm^{-2}}$ is the penetration depth of cosmic rays. \par
The third contribution was the stellar X-ray ionisation rate, calculated at every $R$ and $Z$ as \citep{Igea1999, Bai2009, Jankovic2021},
\begin{multline}
    \zeta_\mathrm{X}(R,Z)=\frac{L_\mathrm{X}}{10^{29}~\mathrm{erg~s}^{-1}}\left( \frac{R}{1~\mathrm{AU}}\right)^{-2.2}(\zeta_1e^{-(\Sigma(R,Z)/\lambda_1)^{c_1}}+\\
    \zeta_2e^{-(\Sigma(R,Z)/\lambda_2)^{c_2}}) ~~~,
\end{multline}
with $L_\mathrm{X}$ being the stellar X-ray luminosity, set to $10^{30}~\mathrm{erg~s}^{-1}$ \citep{Gdel2007}. The parameters $\zeta_1=6\cdot 10^{-12}~\mathrm{s}^{-1}$, $\lambda_1=3.4\cdot 10^{-3}~\mathrm{g~cm}^{-2}$ and $c_1=0.4$ describe the absorption of X-rays and $\zeta_2=10^{-15}~\mathrm{s}^{-1}$, $\lambda_2=1.6~\mathrm{g~cm}^{-2}$ and $c_2=0.65$ are used to treat the scattered X-rays. We note that $\zeta_\mathrm{CR}(R,Z)$ and $\zeta_\mathrm{X}(R,Z)$ do not account for contributions coming through the disc from the opposite side. However, as explained in \cite{Jankovic2021}, these contributions cannot change the resulting non-thermal ionisation rate by more than a factor of 2.

\begin{table}[t]
{\renewcommand{\arraystretch}{1.1}
\caption{Model parameters.}
\begin{tabular}{ll||lcl}
\hhline{=====}
                                                            & 2D+3D              &                           & \multicolumn{1}{r|}{2D}   & 3D        \\ \hline
$M_\star$ {[}$\mathrm{M}_\odot${]}                          & 1.0                & $r_\mathrm{in}$ {[}AU{]}  & \multicolumn{2}{c}{0.05}              \\
$R_\star$ {[}$\mathrm{R}_\odot${]}                          & 2.6                & $r_\mathrm{out}$ {[}AU{]} & \multicolumn{1}{r|}{4}    & 2         \\
$T_\star$ {[}K{]}                                           & 4300               & $\theta$ {[}rad{]}        & \multicolumn{2}{c}{$\pi /2 \pm 0.15$} \\
$\alpha_\mathrm{MRI}$                                       & $10^{-1}$          & $N_\mathrm{r}$            & \multicolumn{1}{r|}{1698} & 512       \\
$\alpha_\mathrm{DZ}$                                        & $10^{-3}$          & $N_\theta$                & \multicolumn{2}{c}{128}               \\
$\dot{M}_\mathrm{init}$ {[}$M_\odot \, \mathrm{yr}^{-1}${]} & $3.6\cdot 10^{-9}$ & $\varphi$ {[}rad{]}       & \multicolumn{1}{l|}{}     & $[0,\pi]$ \\
$f_0$                                                       & $10^{-3}$          & $N_\varphi$               & \multicolumn{1}{l|}{}     & 512  \\
\hline
\end{tabular}
\label{tab:model_params}
}
\end{table}

\subsection{Numerical considerations} \label{sec:numerical}
Due to the substantial computational cost of the models of this work, we adapted the numerical grid to extend radially from $r_\mathrm{in}=0.05\;\mathrm{AU}$ to $r_\mathrm{out}=4\;\mathrm{AU}$ in $N_\mathrm{r}=1696$ logarithmically spaced cells in the 2D models and to $r_\mathrm{out}=2\;\mathrm{AU}$ in $N_\mathrm{r}=512$ logarithmically spaced cells in the 3D model. For both geometries, the polar domain covered a range of $\theta=\pi/2 \pm 0.15$, separated into $N_\theta=128$ linearly spaced cells. For the 3D run, the computational domain included half of the disc in the azimuthal direction ($\varphi=[0, \pi]$) with $N_\varphi=512$ cells and periodic boundary conditions. All other boundary conditions were adopted from \citetalias{Cecil2024b}. \par
As explained in \citetalias{Cecil2026}, the inclusion of the detailed gas opacities from \cite{Malygin2014} necessitates the employment of an under-relaxation scheme in the optically thin disc atmosphere, which has no consequences for the instability mechanism leading to outbursts (for details, we refer to Appendix F of \citetalias{Cecil2026}). Furthermore, a small radial smoothing range was included in the determination of $\alpha$ to avoid strong jumps between adjacent cells, which can cause numerical instability. Similar to an under-relaxation method, we required that the value of $\alpha$ only changes by a percentage $\iota$ of the difference between the values in the present and radially adjacent cell, 
\begin{equation} \label{eq:alpha_smooth}
    \alpha_\mathrm{sm}^{i+1}=\alpha^i+\iota \left(\alpha_\mathrm{true}^{i+1}-\alpha^i\right) \;,
\end{equation}
 \noindent where $\alpha^i$ is the value in the current radial cell, $\alpha_\mathrm{true}^{i+1}$ is the true value, calculated as described in Sect. \ref{sec:meth_viscosity}, in the next radial cell and $\alpha_\mathrm{sm}^{i+1}$ is the final, smoothed value in the neighbouring cell. For the 3D model, we additionally applied a Gaussian blur to the resulting azimuthal $\alpha$ profile with a standard deviation equivalent to five azimuthal cells.

\section{Results}\label{sec:results}

\begin{table}[]
{\renewcommand{\arraystretch}{1.35}
\caption{Model names and configurations}
\centering
\begin{tabular}{c|c|c|c}
\hhline{====}
Model                               & \makecell[c]{$B_\star$\\[-0.01em]{[}kG{]}} & \makecell[c]{$B_\mathrm{disc}$\\[-0.01em]{[}G{]}}   & Geometry\\ \hline
$\texttt{FULL}$      & -                                                           & -                           & 2D       \\
\texttt{STAR1}      & 1                                                                                              & 0      & 2D  \\
\texttt{STAR2} & 2                                                                                               & 0          & 2D    \\
\texttt{STAR1DISC01}      & 1                                                                                               & 0.1   & 2D  \\
\texttt{STAR1DISC013D}      & 1                                                                                               & 0.1   & 3D  \\
\texttt{STAR1DISC01r}      & 1                                                                                               & $\frac{0.1}{r\, [\mathrm{AU}]}$  & 2D   \\      
\texttt{STAR2DISC001}      & 2                                                                                               & 0.01   & 2D  \\      
\texttt{STAR2DISC01}      & 2                                                                                               & 0.1   & 2D  \\      
\texttt{STAR2DISC01r}      & 2                                                                                               & $\frac{0.1}{r\, [\mathrm{AU}]}$   & 2D  \\      
\hline                                                          
\end{tabular}
\label{tab:models}

}
\end{table}

To investigate the impact of MRI activity determined by external magnetic fields and the effects of ambipolar and Ohmic diffusion on the inner disc instability mechanism, we analysed the structure and evolution of the inner disc in seven models that differ in the prescribed magnetic field. As a reference, we compared these models to the \texttt{FULL} model presented in \citetalias{Cecil2026}. Additionally, we described the non-axisymmetric signatures that result from the evolution of an outburst by extending one of the axisymmetric models to 3D. The parameters shared by all simulations are listed in Tab. \ref{tab:model_params}. The names, stellar and disc magnetic field components, as well as the dimensionality of all models are given in Tab. \ref{tab:models}. \texttt{FULL} has the same physical setup, but was constructed with an MRI activation criterion of $T_\mathrm{g}>T_\mathrm{MRI}$ (including a relatively large smoothing range around $T_\mathrm{MRI}$, shown in Eq. 11 of \citetalias{Cecil2026})\footnote{Additionally, \texttt{FULL} included frequency-dependent irradiation from the central star. However, this computationally costly addition does not significantly affect the instability mechanism.}

% \subsection{Bursting behaviour and burst morphologies}
\subsection{A new burst morphology}

\begin{figure}[]
    \centering
         \resizebox{\hsize}{!}{\includegraphics{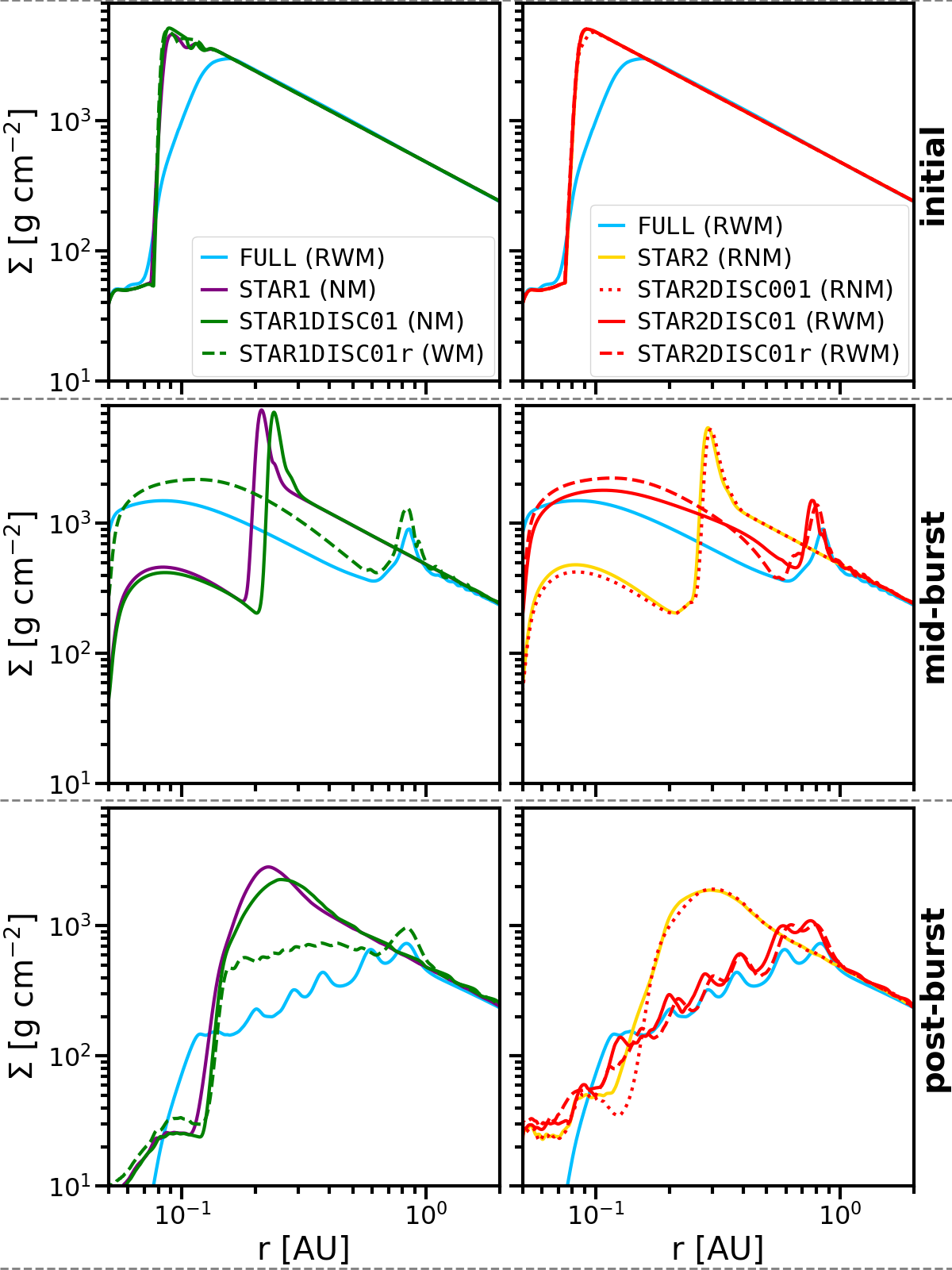}}
    \caption{Surface density profiles of all models listed in Table \ref{tab:models}. The first, second and third rows show the hydrostatic initial models, the state at which the MRI active burst region has reached its largest extent, and the state at the beginning of the quiescent phase after the initial burst, respectively. Models including a 1\;kG stellar magnetic field are shown in the left column, while the right column depicts the models that employ a 2\;kG stellar field. As a reference, the \texttt{FULL} model is included in each panel.}
    \label{fig:sigma_profiles}
\end{figure}

Across all models presented in \citetalias{Cecil2024b} and \citetalias{Cecil2026}, the episodic accretion events exhibited the same robust morphology. The inclusion of MRI activation criteria based on non-ideal MHD effect changes this picture by producing distinctly different burst structures. In fact, a new mode of outbursts manifests in some of our models, which will be referred to here as the narrow burst mode, in contrast to the wide burst mode observed in our previous models. Before we introduce a categorisation of the different burst modes, we first recapitulate the general bursting mechanism. 
\subsubsection{General instability mechanism}
The evolution of the outburst has been extensively described in \citetalias{Cecil2024b} and \cite{Ziampras2026}. In the following, we summarise the key aspects and emphasise the alterations induced by the new MRI activation criteria. \par
The first row of Fig. \ref{fig:sigma_profiles} depicts the initial surface density structures of all 2D models listed in Tab. \ref{tab:models}. There are no significant differences among the setups, enabling us to compare their behaviour during the instability phase in response to the different magnetic field strengths\footnote{The shallower density gradient at the DZIE in the \texttt{FULL} model is caused by the different smoothing description of the models in \citetalias{Cecil2026}.}. \par
As expected, all models immediately enter the burst phase by activating the MRI in the dead zone after initiating the radiation hydrodynamic simulation. 
The runaway heating effect, caused by the suddenly enhanced viscous energy dissipation and trapping of heat near the midplane in the optically thick inner dead zone, leads to a quick expansion of the MRI active region in both radial and vertical directions. The expansion is led by a heating front across which the MRI becomes active. Due to the substantial difference in angular momentum transport efficiency between the regions in front and behind the heating front, a density spike forms and propagates outward, ahead of the front. As long as the conditions for MRI activation can be met in this high-density region, the MRI active zone continues to expand. The second row of Fig. \ref{fig:sigma_profiles} shows the surface densities at the respective moments when the radially outward-moving heating front stalls (mid-burst) in each model. The highly turbulent material in the enlarged MRI active regions is quickly flushed onto the star, allowing the inner disc to cool again. The heating front is then travelling back towards the star as a cooling front, continuously shutting off the MRI activity. In all our previous models (including \texttt{FULL}), this mechanism immediately repeated itself by reigniting the MRI when the cooling front entered the innermost, stellar-heating-dominated region. During this reflaring of the outburst, the heating front stalls at smaller distances due to the reduced amount of material in the inner disc following the previous flare. Only after several reflares have occurred does the burst cycle end, and the disc enters the quiescent phase. In the models of this work, the reflaring behaviour only manifests in simulations with $B_\star=2\,\mathrm{kG}$ (right column of Fig. \ref{fig:sigma_profiles}), while the burst phase ends after the first flare in models with  $B_\star=1\,\mathrm{kG}$.

\subsubsection{Burst mode categorisation}

\begin{table}[t]
\centering
{\renewcommand{\arraystretch}{1.2}
\caption{Burst modes appearing in our models}
\begin{tabular}{c|c|c}
\hhline{===}
\multicolumn{1}{l|}{} & Reflaring                                                                                                                                               & {\renewcommand{\arraystretch}{1.0}\begin{tabular}[c]{@{}c@{}}No\\ reflaring\end{tabular} }                                                         \\ \cline{1-3} 
Wide                  & \begin{tabular}[c]{@{}c@{}}\texttt{FULL}\\ \texttt{STAR2DISC01}\\ \texttt{STAR2DISC01r}\end{tabular} & \texttt{STAR1DISC01r}                                                                          \\ \cline{1-3} 
Narrow                & \begin{tabular}[c]{@{}c@{}}\texttt{STAR2}\\ \texttt{STAR2DISC001}\\

\end{tabular}                                        & \begin{tabular}[c]{@{}c@{}}\texttt{STAR1}\\ \texttt{STAR1DISC01}\end{tabular} \\
\hline
\end{tabular}

\label{tab:burst_modes}
}
\end{table}

Fig. \ref{fig:sigma_profiles} shows an apparently discrete distinction between two modes of bursts. In the wide burst mode, the ionisation front is able to travel outwards to the same radial location as in our previous models. 
On the other hand, if the considered magnetic field only consists of the stellar component, or the disc component is too weak, the outward movement of the ionisation front is stopped at much smaller distances. Even though there is a massive density accumulation just ahead of the front, the conditions for MRI activity cannot be met, establishing the narrow burst mode. The underlying reason for this dichotomy is explored in Sect.~\ref{sec:res_dichotomy}.\par
The different manifestations of the burst mechanism appearing in our models can be categorised into four bursting modes, solely determined by the magnetic field configuration:
\begin{itemize}
    \item Narrow mode (NM)
    \item Reflaring narrow mode (RNM)
    \item Wide mode (WM)
    \item Reflaring wide mode (RWM)
\end{itemize}
The categorisation of the models analysed in this work is summarised in Tab.~\ref{tab:burst_modes}. \par
The burst mode has a crucial influence on the post-burst states, shown in the third row of Fig.~\ref{fig:sigma_profiles}. In the RWM, which occurs in models with both a strong stellar and disc magnetic field component, the post-burst disc structure closely resembles that in models with a simple $T_\mathrm{MRI}$ description. The initial propagation of the heating front reaches the same distance from the star and multiple density and pressure bumps are placed by the reflares throughout the inner disc. Without reflares, in the WM, the outermost density maximum still forms at the same distance, but the inner disc shows an almost flat surface density profile towards the DZIE. On the other hand, in the NM and RNM, only one massive density bump at small radii ($\lesssim 0.3$ AU) results from the outburst, with the surface density structure appearing similar to the initial configuration. The reflares in the RNM do not significantly affect the post-burst state. \par

\begin{figure}[t]
    \centering
         \resizebox{\hsize}{!}{\includegraphics{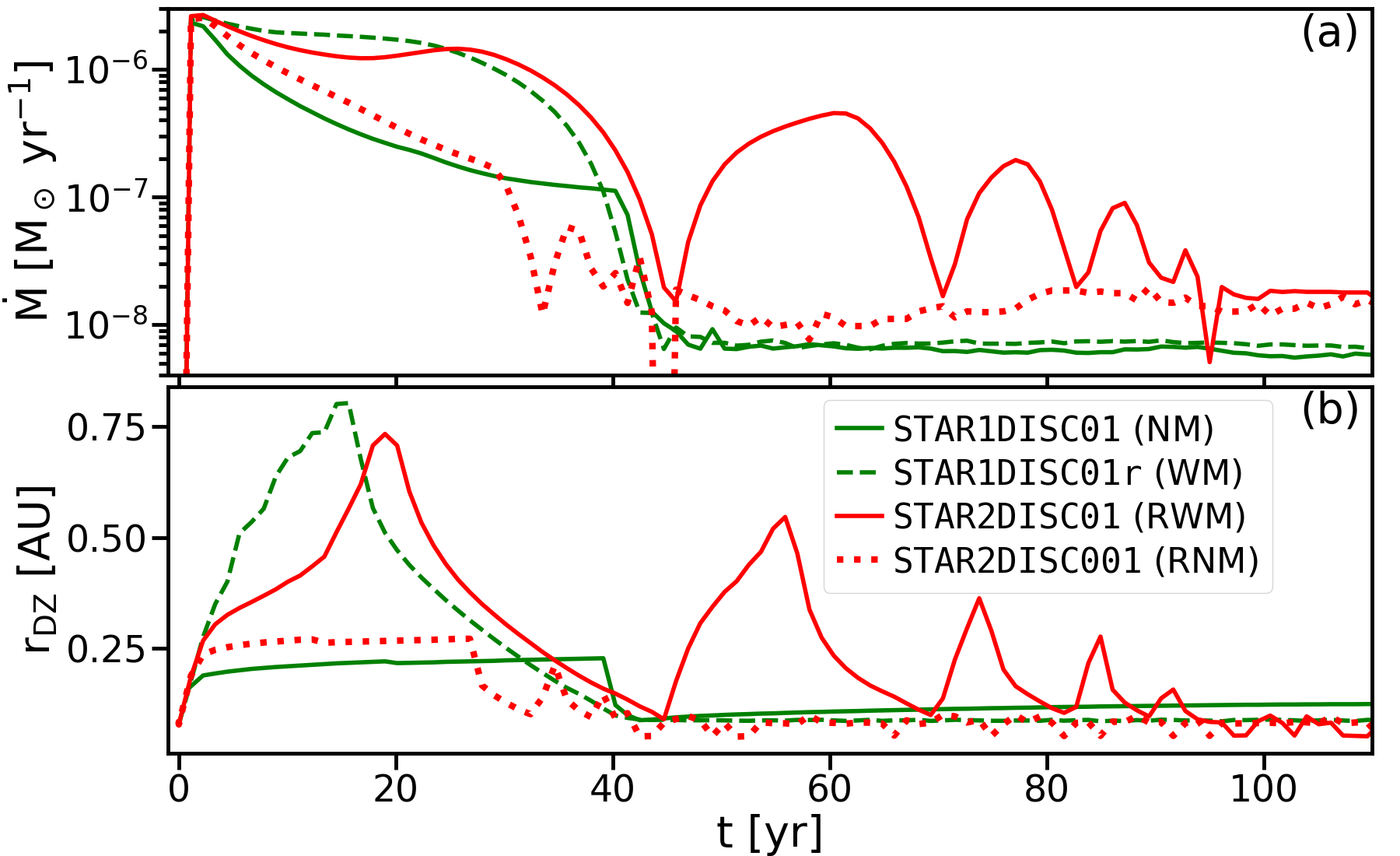}}
    \caption{Accretion rates (panel a) and radii of the DZIE at the midplane (panel b) of models displaying the four different burst modes.}
    \label{fig:accr_rdz_morphos}
\end{figure}

\begin{figure*}[h]
    \centering
         \resizebox{\hsize}{!}{\includegraphics{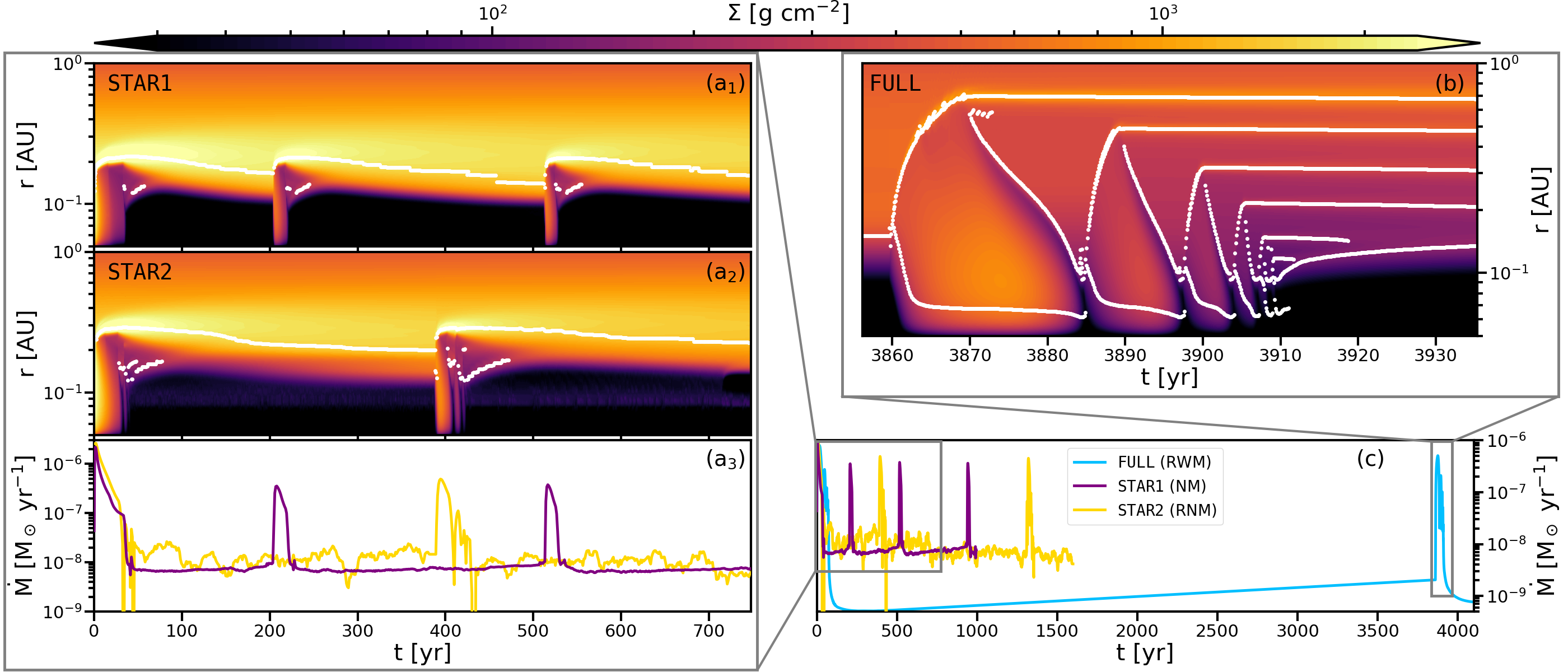}}
    \caption{Comparison of the accretion rates and surface density evolutions of the \texttt{FULL} model (with the $T_\mathrm{MRI}$ MRI activity criterion) and the two models including only the stellar magnetic field component, \texttt{STAR1} and \texttt{STAR2}. Panel c shows the accretion rates of all three models. Panel ($\mathrm{a}_3$) depicts a zoom-in to the time-frame that captures the bursting-timescale of the \texttt{STAR1} and \texttt{STAR2} models, while panels ($\mathrm{a}_1$) and ($\mathrm{a}_2$) display their respective surface density evolution. The space-time diagram of the outburst occurring in the \texttt{FULL} model is provided in panel b. The white indicators in panels ($\mathrm{a}_1$), ($\mathrm{a}_2$) and b mark the positions of local pressure maxima.  }
    \label{fig:radius_time}
\end{figure*}
Panel a of Fig. \ref{fig:accr_rdz_morphos} shows the morphologies of all four burst modes in the context of the accretion rate. We stress that, due to our choice of initial conditions, the absolute and relative values of the accretion rates may not necessarily match those of a "naturally" occurring burst ignited after full evolution of the quiescent phase between accretion events. However, the shapes of the accretion rate curves are expected to remain the same for each burst cycle in the respective mode (as analysed in \citetalias{Cecil2024b} and \citetalias{Cecil2026} for the RWM and indicated in Fig. \ref{fig:radius_time} for the NM and RNM). The corresponding locations of the DZIE at the midplane are depicted in panel b. In the NM and RNM, the ionisation front stalls before 0.3~AU. Consequently, the amount of mass available to be flushed onto the star is much smaller than in the WM. This leads to a quicker decrease in the accretion rate after the initial peak. On the other hand, the curves in panel a for the WM and the first flare of the RWM are almost equal. The small valley in the profile of \texttt{STAR2DISC01} at a time of around 15~yr is due to the onset of hydrodynamic instability of the region around the ionisation front (Sec. \ref{sec:res_dichotomy}). The reflaring behaviour of \texttt{STAR2DISC01} can be attributed to the stronger magnetic field close to the star, which decreases the $\beta$ parameter enough behind the retreating cooling front to establish similar conditions to our previous models with a $T_\mathrm{MRI}$ transition, all of which showed reflares.

\subsection{Accretion history and evolution of pressure maxima}
The models \texttt{STAR1} and \texttt{STAR2} were evolved long enough to capture multiple episodic accretion events. Panel c of Fig. \ref{fig:radius_time} shows a comparison between their accretion rate histories and that of \texttt{FULL}. The frequency of outbursts depends on the occurrence of reflares and can be understood as a function of the maximum expansion of the MRI active region during the instability phase. In the RWM, the burst removes material from the inner disc from radii of up to 0.8\;AU as long as reflares can be ignited. Therefore, the refilling of the inner disc by material being accreted from larger radii requires significantly more time before a new burst cycle can be initiated compared to the RNM, where the progression of the heating front is already stopped before reaching 0.3\;AU. The quiescent accretion rates in \texttt{STAR1} and \texttt{STAR2} are higher than in \texttt{FULL} and tend to decrease with each accretion event (especially in the case of \texttt{STAR2}). We explore this behaviour further in Sect. \ref{sec:long_term}. \par
Panel ($\mathrm{a_3}$) captures the bursting timeframe of \texttt{STAR1} and \texttt{STAR2}, during which the RNM reflares are clearly visible. The values of the accretion rate curve of \texttt{STAR2} displayed in panels ($\mathrm{a_3}$) and c have been averaged over 1500 inner orbits (corresponding to $\sim\!\!17$\;yr) during the quiescent phases (i.e. the accretion rates during the burst phases have been left unchanged) since they showed significant stochasticity, which is still apparent in the averaged values. The underlying reason will be explored in Sec. \ref{sec:structure}. The burst frequency is highest in \texttt{STAR1} since the MRI active region cannot expand past $\sim\!\!0.2$\;AU and reflares are absent, retaining more mass in the inner disc and allowing another outburst to occur sooner. \par
A crucial aspect of the models presented in \citetalias{Cecil2024b} was the analysis of the destruction and formation of pressure maxima throughout the inner disc by bursts in the RWM. We showed that the stable pressure maximum at the DZIE during the quiescent phase \citep[which was recognised as an effective pebble and planet migration trap, e.g.,][]{Flock2019} is disrupted by the instability mechanism. However, the bursting process produces multiple additional pressure bumps whenever the heating front reaches its maximum extent during each flare and reflare. This phenomenon is depicted again in panel b of Fig. \ref{fig:radius_time} for the example of the \texttt{FULL} model. Only after the burst cycle has concluded does the stable pebble trap at the DZIE reappear (pressure bump closest to the star after $t\sim 3910$\;yr in panel b). This picture shifts significantly in the NM and RNM. \par
The evolution of the surface density and location of pressure maxima is very similar for the models \texttt{STAR1} and \texttt{STAR2}, as shown in panels ($\mathrm{a_1}$) and ($\mathrm{a_2}$) of Fig. \ref{fig:radius_time}. However, they are remarkably different from the RWM shown in panel b: There is only one significant pressure bump present at any time, which is periodically pushed outward by the burst. It corresponds to the density maximum in the post-burst states of models with the NM or RNM shown in the bottom row of Fig. \ref{fig:sigma_profiles}. During quiescence, this peak slowly moves back towards the star, increasing the density at small radii until the conditions for MRI activation are satisfied again and a new burst cycle is initiated. \par

\begin{figure*}[t]
    \centering
         \resizebox{\hsize}{!}{\includegraphics{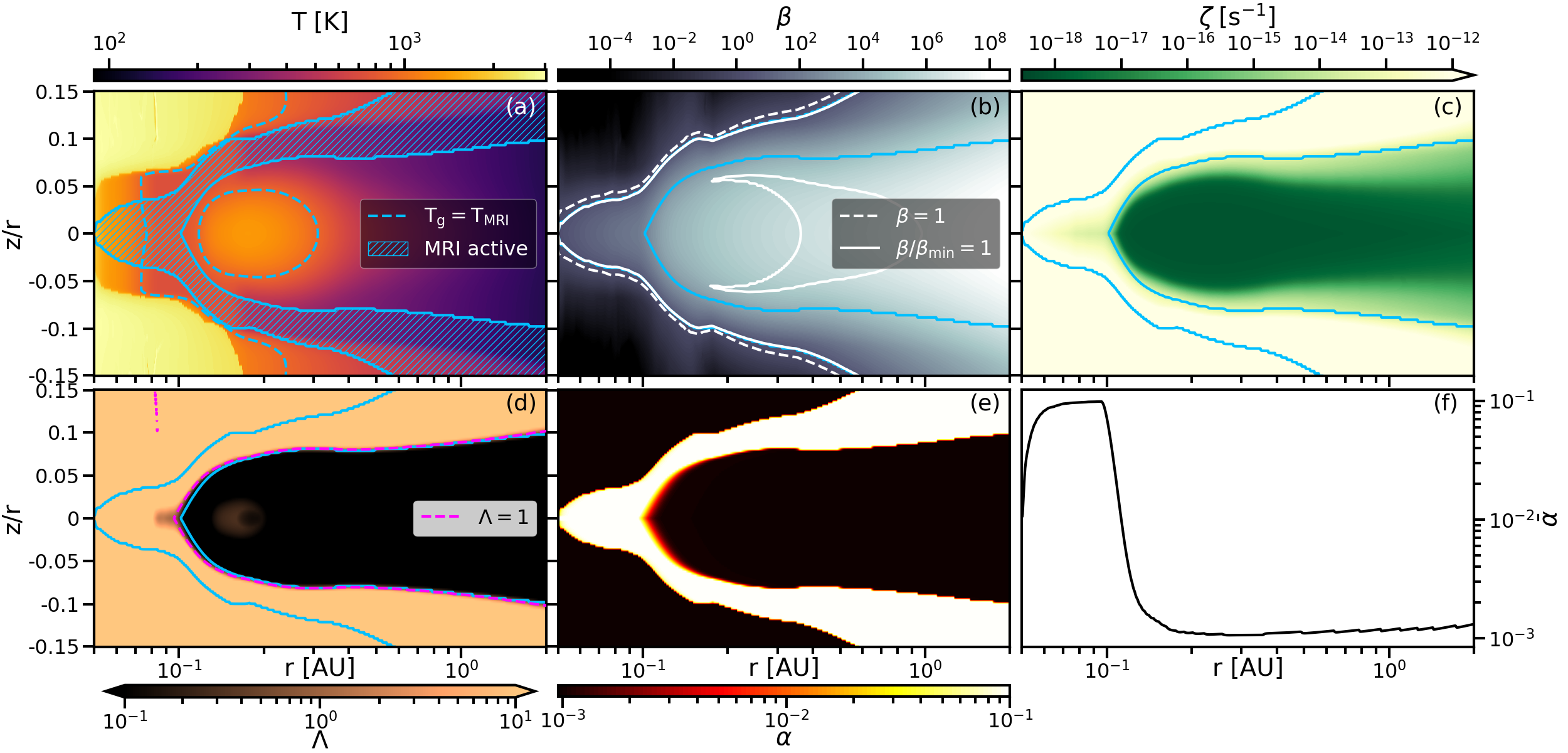}}
    \caption{ Two-dimensional heat maps of several quantities of the \texttt{STAR1} model in the quiescent phase. Panels a, b and c show the temperature, plasma $\beta$ and non-thermal ionisation rate, respectively, and panels d, e and f display the Ohmic Elsässer number, local and vertically integrated $\alpha$, respectively. The MRI active regions are located between the blue contour lines, as indicated by the hatched area in panel a. The dashed blue line in panel a and the solid white line in panel b indicate the MRI transitions according to the $T_\mathrm{MRI}$ and $\beta>\beta_\mathrm{min}$ criterion, respectively. The dashed white line in panel b marks the contour of equal magnetic and thermal pressure ($\beta=1$). The pink-dashed line in panel d marks the contour of the Ohmic Elsässer number equal to unity.}
    \label{fig:structure}
\end{figure*}

Practically, we can understand the difference in the context of pressure maxima locations between the (R)WM and the (R)NM in the following way: While pressure bumps are destroyed and (re)created by the burst mechanism in the (R)WM, the single pressure maximum at the DZIE is just pushed outwards before slowly retreating again, never staying static, in the (R)NM.

\subsection{The structure of the inner disc} \label{sec:structure}
The MRI activity criteria described by Eqs. \ref{eq:MRIcrit1} and \ref{eq:MRIcrit2} can result in significant differences in the inner disc's architecture compared to a simple $T_\mathrm{MRI}$ threshold. The main contributors to the determination of MRI activity, as well as its major effects in the context of our model setup, are shown in Fig. \ref{fig:structure} for the example of the \texttt{STAR1} model in quiescence. \par
Panel a provides a direct comparison between the MRI activity resulting from our current criteria (blue hatched region) and the $T_\mathrm{MRI}=900$\;K condition used in our previous work. In the thin disc atmosphere, the MRI is mostly suppressed by ambipolar diffusion. The contour for the corresponding $\beta_\mathrm{min}$ value is marked in panel b and is almost equivalent to the $\beta=1$ surface. Since this model does not include a disc magnetic field component, the plasma $\beta$ becomes very large as the dipolar stellar magnetic field strength quickly decreases with radius. \par
According to the $T_\mathrm{MRI}$ condition, the structure shown in Fig. \ref{fig:structure} should already have entered the burst phase, as indicated by the blue dashed line in panel a. However, the MRI in the optically thick region around the midplane remains quenched by Ohmic diffusion, as apparent in the map of the Ohmic Elsässer number, shown in panel d. Panel c displays the non-thermal ionisation rate, which reaches values larger than $10^{-12}\;\mathrm{s^{-1}}$ in the optically thin regions, predominantly due to X-ray ionisation. This leads to a MRI active layer between the Ohmic dead zone around the midplane and the MRI inactive region due to ambipolar diffusion in the low--$\beta$ regions. \par
The resulting distribution of local $\alpha$ values is shown in panel e. The abruptness of the MRI transition, as defined by the criteria given by Eqs. \ref{eq:MRIcrit1} and \ref{eq:MRIcrit2} has been slightly smoothed by applying Eq. \ref{eq:alpha_smooth}. Panel f depicts the radial profile of the pressure-averaged, vertically integrated viscosity parameter, calculated as, 
\begin{equation} \label{eq:alphabar}
    \bar{\alpha}=\frac{\int_{-\infty}^{\infty}\alpha P_\mathrm{g}\;dZ}{\int_{-\infty}^{\infty}P_\mathrm{g}\;dZ} \;,
\end{equation}
\noindent where $Z$ is the cylindrical height. Panels e and d indicate that, close to the inner boundary, the MRI is almost completely quenched due to the high magnetic field strength and low density, which push $\beta$ below $\beta_\mathrm{min}$ near the midplane. In models with a stronger stellar magnetic field, the MRI can be suppressed even at the midplane during quiescent phases. This can lead to stochastic behaviour of the accretion rate on short timescales due to flickering of the MRI activity at the midplane near the inner boundary, as small amounts of material constantly accumulate and drain in these rapidly evolving regions. This stochasticity is indicated in the accretion rate of \texttt{STAR2} in Fig. \ref{fig:radius_time}. \par
In models including a disc magnetic field component, the structure of the MRI active regions remains very similar to what is shown in Fig. \ref{fig:structure}. The most significant differences arise due to smaller values of $\beta$, especially in regions where the stellar magnetic field component becomes subdominant. This leads to a shift of the MRI active layer closer towards the midplane, decreasing the vertical extent of the dead zone. Consequently, $\bar\alpha$ can increase appreciably above $\alpha_\mathrm{DZ}$ at larger radii. We refer to Fig. \ref{fig:extrapolation} for the structure of the MRI active regions in the mid-burst state of the NM and RWM, respectively. 

\subsection{The dichotomy of narrow and wide bursts} \label{sec:res_dichotomy}

In \citetalias{Cecil2026}, we investigated the susceptibility of the burst features to the RWI and found that the instability is indeed expected to develop at the density bumps placed by the burst cycle (we continue this investigation in Sect. \ref{Res_sec:3D} in the context of the model \texttt{STAR1DISC013D}). However, in our axisymmetric models, RWI cannot manifest, unlike potential Rayleigh instability, which primarily describes unstable motion in the radial direction. We concluded in \citetalias{Cecil2026} that during the burst, the outward-moving density spike might also become Rayleigh unstable, based on the criterion $\kappa_\mathrm{ef}^2<0$, where $\kappa_\mathrm{ef}$ is the epicyclic frequency. Since $\kappa_\mathrm{ef}^2$ only represents the gradient of the specific angular momentum, this criterion does not consider the stabilising or destabilising contribution by a potential entropy gradient. However, since the entropy in our models cannot be assumed to be constant with radius or height, a more detailed investigation of the dynamical stability requires the assessment of the Solberg--H{\o}iland criterion, $C_\mathrm{SH}<0$, as a generalised version of the Rayleigh criterion \citep[e.g.,][]{Yang2010}. We explore the functional form of $C_\mathrm{SH}$ and its limitations in the context of our models in Appendix \ref{sec:app_solberg}.

\begin{figure}[]
    \centering
         \resizebox{\hsize}{!}{\includegraphics{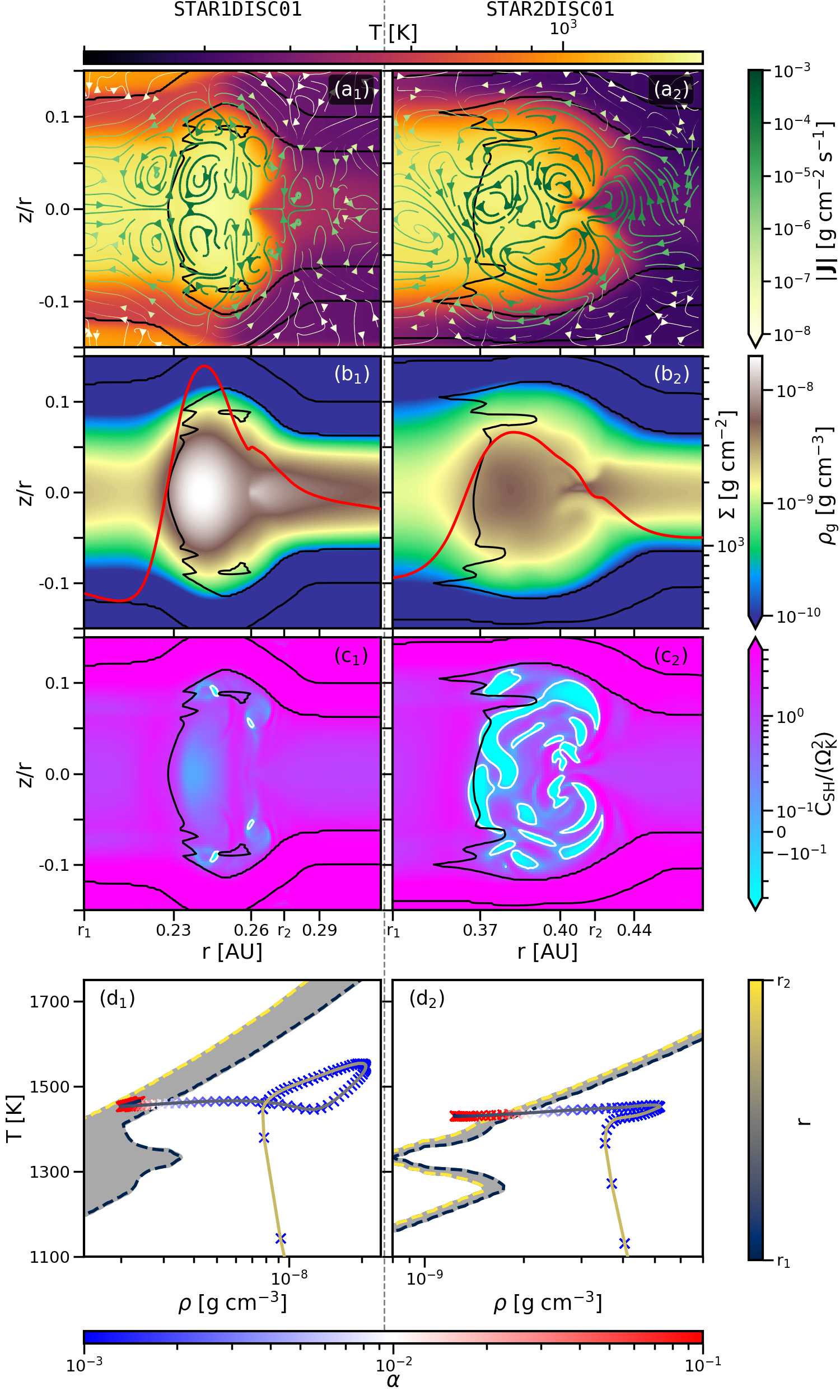}}
    \caption{Structure and behaviour of the region around the ionisation front in the case of a narrow (left column, model \texttt{STAR1DISC01}) and wide burst (right column, model \texttt{STAR2DISC01}). The first row shows the temperature structure and the mass flux density field. The arrows are colour-coded according to the local magnitude of the flux density $|\vec{\mathrm{J}}|$. The second row depicts the density maps and the surface density profile (red line). In the third row, the distributions of the normalised values of the Solberg--H{\o}iland condition are illustrated. The transitions from positive to negative values are marked as white contour lines. The black contours in the first three rows correspond to transitions from MRI active to inactive. The bottom row displays the MRI transition region for radii between $r_1$ and $r_2$ (marked on the x-axis of the third-row panels) in the $\rho-T$ plane, shaded in grey. The transitions at $r_1$ and $r_2$ are specifically marked with blue and yellow dashed lines, respectively. The panels $\mathrm{d_1}$ and $\mathrm{d_2}$ also include the radial distribution of midplane density and temperature values for the respective model between $r_1$ and $r_2$. While the colour of the line corresponds to the radius $r$, the crosses along the line are colour-coded according to the midplane value of $\alpha$.  }
    \label{fig:SH}
\end{figure}

Figure \ref{fig:SH} compares the conditions around the ionisation front during a burst between two models displaying the narrow (\texttt{STAR1DISC01}, left column) and wide (\texttt{STAR2DISC01}, right column) burst mode. We chose these two models because their disc magnetic field components are equal. The respective radial ranges include the regions where the ionisation front (vertical black contour through the midplane in the upper three rows) is expected to stall. Due to the stronger stellar magnetic field in \texttt{STAR2DISC01}, this region lies at a larger radius. A comparison of the maximum radii up to which the MRI can be activated between the two models is provided in Appendix \ref{sec:app_betaradius}. In the \texttt{STAR1DISC01} model, the movement of the ionisation front is indeed halted, while for the \texttt{STAR2DISC01} model, the front continues to progress outwards. Panels ($\mathrm{d_1}$) and ($\mathrm{d_2}$) show a clear spike in midplane density for both models (as also apparent in panels $\mathrm{b_1}$ and $\mathrm{b_2}$). However, this density accumulation alone is insufficient to activate the MRI, since the temperature within it remains well below the required value. For the ionisation front to progress, either the temperature needs to increase much further, or the density needs to decrease while keeping the same temperature within the density spike. Neither of these possibilities is realisable in a dynamically stable environment in thermal equilibrium, especially since a further increase in temperature is strongly impeded by the thermostat effect arising from dust sublimation. Panel ($\mathrm{c_1}$) indicates that the condition for dynamic instability according to Eq. \ref{eq:SH1} is not fulfilled anywhere in the depicted region in the case of \texttt{STAR1DISC01}. The mass flux density field illustrated in panel ($\mathrm{a_1}$) shows that small vortices have still formed around the ionisation front, which, however, can be explained by the disc material responding to the strong pressure gradients. These disturbances are mostly symmetric between the hemispheres and are small enough not to significantly affect the density and temperature structure around the midplane. Therefore, the conditions just outside the current location of the ionisation front cannot be pushed into the MRI active regime, and the ionisation front stalls. \par
In contrast, panel ($\mathrm{c_2}$) shows that $C_\mathrm{SH}/(\Omega_\mathrm{K}^2)$ decreases well below zero in large regions ahead of the ionisation front in the model \texttt{STAR2DISC01}. This results in strong, asymmetric disturbances and convection-like motion of the disc material, as evident in panel ($\mathrm{a_2}$). As a consequence, the density spike becomes smeared out (as can be judged by comparing the surface density profiles in panels $\mathrm{b_2}$ and $\mathrm{b_1}$), and the turbulent motion can eventually establish MRI-favourable conditions in the midplane, allowing the ionisation front to progress further. This is additionally aided by the MRI activation criterion being shifted to lower temperatures in \texttt{STAR2DISC01} compared to \texttt{STAR1DISC01} due to the stronger stellar magnetic field, despite the greater distance from the star (panels $\mathrm{d_1}$ and $\mathrm{d_2}$). \par 
Even though the density and pressure gradients are much stronger in the case of \texttt{STAR1DISC01}, the region around the ionisation front remains dynamically stable (except for RWI in non-axisymmetry, see Sec. \ref{Res_sec:3D}). This is mainly due to the strong radial dependence of $C_\mathrm{SH}$: All three components ($N_\mathrm{R}^2,~N_\mathrm{Z}^2$ and $\kappa_\mathrm{ef}^2$) typically decrease with distance from the star, facilitating the potential fulfilment of Eq. \ref{eq:SH1}. Additionally, the viscous timescale ($t_\mathrm{\nu}=r^2/\nu$) is larger for greater distances, making the suppression of large-scale turbulent motion more difficult at larger radii. \par
Panel ($\mathrm{d_2}$) also shows a clear kink in the MRI transition line at low densities. It originates from the sublimation of dust and the consequential cessation of thermionic emission at higher temperatures. We elaborate on the influence of this effect in Appendix \ref{sec:app_thermionic}. \par
We conclude that the dichotomy of the burst modes is established due to the onset of dynamic instability of the region around the ionisation front at a certain distance from the star. If the magnetic field is strong enough to drive the ionisation front to this distance, the subsequent progression is governed by the dynamic instability, which enables the wide burst mode. At large distances, the propagation is further aided by the contribution of thermionic emission ahead of the ionisation front. Otherwise, the front stalls before the accompanying density accumulation can become unstable according to the Solberg--H{\o}iland criterion (and temperatures are above the dust sublimation threshold ahead of the front), only allowing for the narrow burst mode.

\begin{figure*}[]
    \centering
         \resizebox{\hsize}{!}{\includegraphics{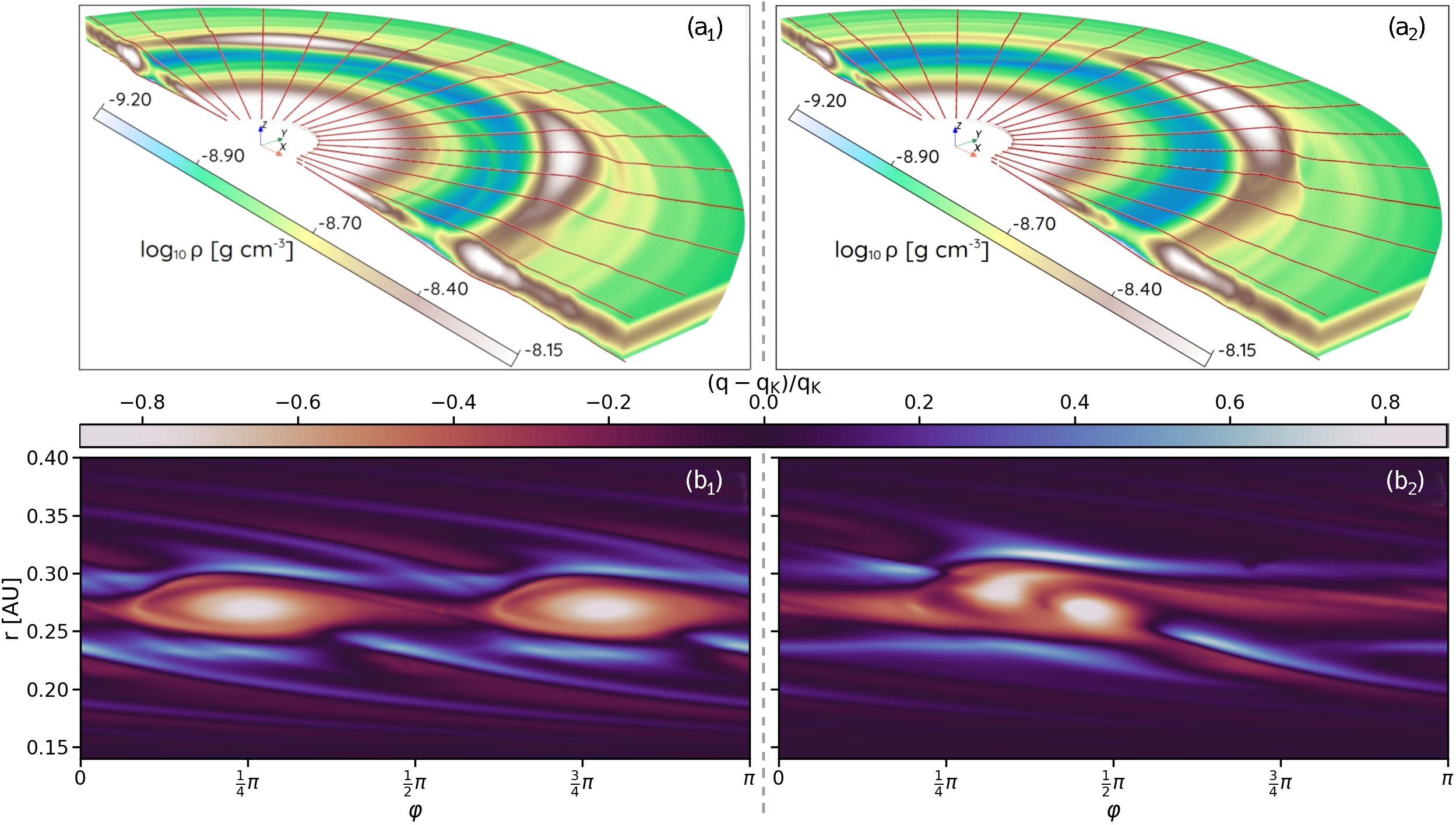}}
    \caption{Emergence of vortices in the \texttt{STAR1DISC013D} model. The left column shows the disc's state just after the RWI has set in and formed two distinct vortices. The right column represents the disc at a time of about 20 local orbits (at the location of the vortices) later. The upper row shows 3D renderings of the disc's mass density out to 0.5 AU in a Cartesian coordinate system. The height of the uppermost rendered surface corresponds to $1.7\,H$. The red lines trace the surface's shape at different azimuths. The vertical density structure is visible at the azimuthal boundaries. The lower row displays the difference in vertically integrated vortensity between the model and the local Keplerian value in the radial region around the vortices.  }
    \label{fig:3D_vortices}
\end{figure*}

\subsection{Vortex formation in 3D} \label{Res_sec:3D}
In a differentially rotating disc, the RWI develops by trapping Rossby waves within extrema of potential vorticity (or vortensity), which can occur at significant density peaks. The RWI gives rise to strong vortices, which we might expect to observe in non-axisymmetric settings at the strong density features produced by outbursts. \par
While we showed in Sect. \ref{sec:res_dichotomy} that the density spike produced by the (R)NM is not susceptible to Solberg--H{\o}iland-type instability, it is not stable with respect to RWI. To confirm this claim and analyse the non-axisymmetric evolution of this burst feature, we conducted a 3D simulation with the same setup as \texttt{STAR1DISC01}. \par
The activation of the MRI in the dead zone of the initial model of \texttt{STAR1DISC013D}, as well as the propagation of the heating front, remained mostly axisymmetric and equivalent to the 2D model. Differences appeared as soon as the expansion of the MRI active region stalled, and the stage shown in Fig. \ref{fig:SH} was established. Within a few local orbits at the location of the midplane heating front at 0.24~AU, the density spike broke into two distinct vortices. A snapshot of this stage is shown in the left column of Fig. \ref{fig:3D_vortices}. The 3D rendering of the density structure in panel ($\mathrm{a_1}$) reveals clear disc material aggregations at the vortex locations. The red lines, drawn at a height of about 1.7 times the disc scale height $H$, indicate that the scale height has a maximum at the vortices, before dropping off at larger radii. Additional bumps in the scale height profiles are visible along the small spiral structures induced by the strong vortices. Panel ($\mathrm{b_1}$) depicts the residual vertically integrated vortensity map of the radial region around the heating front in units of the Keplerian vortensity. We calculated the vertically integrated vortensity according to the description given in \cite{Masset2016} for a 3D disc setup,
\begin{equation}
    q=\left[\int_{-\infty}^{+\infty} \frac{\rho_\mathrm{g}}{(\vec{\nabla}\times\vec{v})_\mathrm{z}}\;\mathrm{d}Z\right]^{-1} \;,
\end{equation}
\noindent where the cylindrically vertical component of the vorticity is,
\begin{equation}
    (\vec{\nabla}\times\vec{v})_\mathrm{z}= (\vec{\nabla}\times\vec{v})_\mathrm{r} \, \mathrm{cos}\theta - (\vec{\nabla}\times\vec{v})_\mathrm{\theta}\, \mathrm{sin}\theta \;,
\end{equation}
\noindent and the Keplerian vorticity is $(\vec{\nabla}\times v_\mathrm{K})=\Omega_\mathrm{K}/2$. The map shows two clear minima in vortensity, indicating the formation of anticyclonic vortices. Since our azimuthal domain is constrained to $[0,\pi]$, this corresponds to an $m=4$ mode of the RWI. The vortices stretch over almost a quarter of the disc each, corresponding to an extent of 0.4 AU in the azimuthal direction at the radius of the minima. The vortices also excite spiral features in the vortensity and density distributions. About 2.8~yr after the two vortices formed, they merge into a single massive vortex, manifesting the $m=2$ mode. The vortex remains at the position just ahead of the heating front until it is dissolved approximately 150 local orbits ($\sim\!\!20$ yr) after the merger event. During its lifetime, the vortex undergoes several regeneration phases, during which the vortensity minimum deepens again, recompacting the dissolving vortex and extending its lifetime. This results from the competition between the timescale of the density build-up ahead of the ionisation front by angular momentum redistribution and the timescale of RWI saturation, which tends to flatten the density profile. Only after the density behind the heating front has decreased to a degree that the density spike cannot be built back to produce a deep enough vortensity minimum does the vortex diffuse completely. Due to the high computational cost of the 3D model, the simulation was concluded after the disc no longer showed any non-axisymmetric features, but before the MRI active burst region could retreat back towards the star.

\section{Discussion}\label{sec:discussion}
Our models show that including MRI activation descriptions based on non-ideal MHD effects, combined with imposed large-scale magnetic fields, leads to significant changes in the structure of MRI active regions in the inner disc and in the evolution and morphologies of episodic accretion events. In the following, we discuss the potential implications of these findings, compare them to established models and lay out the limitations and future prospects of this work.

\subsection{Inner disc density alteration in the (R)WM and (R)NM} \label{sec:dis_density}
The results of \citetalias{Cecil2024b} made clear that an accretion outburst triggered by the activation of the MRI in the inner dead zone substantially rearranges the structure of the affected burst region, which extends to up to 1~AU in the models therein. However, our simulations show that the robust morphology of those bursts actually represents only one of four different possible burst modes. The newly discovered (R)NM appearing in our models does not have the same significant implications for the inner disc structure. While the pebble- and planet-trapping pressure maximum at the inner edge of the dead zone \citep[e.g.,][]{Flock2019, Paardekooper2022} is destroyed during the RWM burst and reappears in quiescence, it is radially oscillating with an amplitude of approximately 0.1~AU in the case of the (R)NM (Fig. \ref{fig:radius_time}). This implies that the pressure maximum in the (R)NM is never static over extended periods. As soon as it has moved back towards where the stellar magnetic field can sufficiently decrease $\beta$ to activate the MRI in the dead zone, a new burst cycle is initiated, pushing the pressure maximum outwards again. Consequently, while the bump at the DZIE might still act as an efficient pebble trap during quiescence in the (R)WM, it might lose this property in the (R)NM. This could inhibit large amounts of material from accumulating near the DZIE and pose a significant difficulty for in-situ planet formation \citep[e.g.,][]{Hu2026}.\par
Furthermore, the post-burst density structure in the (R)NM does not show a large region with a positive surface density gradient within 1~AU, as was the case for the RWM (Fig. \ref{fig:sigma_profiles}). Such a profile has been suggested as being favourable for enabling convergent migration, a possible pathway for the formation of the terrestrial planets of the Solar System \citep[e.g.,][]{Ogihara2018,Broz2021, Woo2023, Nesvorn2025}. Instead, the hydrostatic structure of our initial models \citep[as presented in detail in][]{Flock2019} is approximately sustained while the DZIE continuously shifts due to burst cycles throughout the evolution of the (R)NM. Simulations of the NM regime, including pebble and dust drift, as well as migrating planets in the inner disc, are required to study the effect of the (R)NM on the torques on planets and trapping of solids close to the star. 

\subsection{Allowed burst modes given by magnetic field strengths}
Given that the mode of the inner disc instability mechanism can have different crucial effects on the structure and evolution of the regions where most exoplanets are detected today, it is essential to differentiate between the modes. As our models indicate, the differentiation is made by the radial profile of the magnetic field strength and the resulting distribution of $\beta$ values throughout the inner disc. In the following, we refer to several examples of observationally and theoretically derived field strengths in the inner disc and relate them to the potential bursting morphology predicted by our models. \par  
Paleomagnetic studies of Solar System bodies point to relatively strong magnetic fields with strengths of >0.4~G even beyond 1~AU in the solar nebula. Taking the large uncertainties of these measurements into account \citep[e.g.,][]{Mansbach2024} and assuming that the magnetic field strength does not decrease towards smaller radii, these records indicate that the field in the protosolar nebula was stronger than the configurations we adopt in our models. Hence, it can be inferred that the PPD of the Solar System could have been subject to the (R)WM of MRI-triggered outbursts, thereby maintaining the possibility of density structures that allow convergent migration (Sect. \ref{sec:dis_density}). \par
Theoretical investigations of magnetic field strengths have mostly relied on partial conservation of magnetic flux during protostellar collapse \citep[e.g.,][]{Masson2015, Vaytet2018, Machida2019}. This can yield hot, strongly magnetised inner discs ($10^2 - 10^3$ G) after the second collapse, as found by \cite{Ahmad2025}, who also argue that the MRI might operate there. These high field strengths could also favour bursts in the (R)WM, although applying our framework to very young, deeply embedded protostars might be difficult. \par
For early Class II discs, which our models match more closely, expected magnetic field configurations and strengths have been calculated, for instance, by \citet{Flock2017a}, \citet{Dudorov2022}, \citet{Khaibrakhmanov2024}, and, most recently, by \citet{Steiner2025}, who established steady-state solutions. In a vertical net-flux model, \citet{Flock2017a} found a radially decreasing (with $\sim\! r^{-3/2}$) mean magnetic field strength with values well above 0.1~G at 1 AU, placing these models within the (R)WM of bursts as well. The calculations shown in \citet{Dudorov2022} and \citet{Khaibrakhmanov2024} exhibit similar profiles but also include a stellar field component and indicate a significant dependence on the dust size assumed in their simulations. On the other hand, the steady-state models by \citet{Steiner2025} result in a dominant stellar magnetic field at distances of up to 1~AU from the central star, outside of which the field strength saturates at values below $10^{-2}$~G while inside the dead zone. They also showed that these specific findings are roughly independent of their choice of $\dot M_\mathrm{init}$ and, by extension, of the $\Sigma$ profile. Hence, we can conclude that such a magnetic field configuration would only allow for the (R)NM.

\subsection{Comparison to prior models}
In the context of MRI activated accretion outburst morphologies, we aim to connect with the methods used by \citet{Vorobyov2020b}, \citet{Kadam2022}, and \citet{Das2025}, who conducted vertically integrated, 2D simulations in the thin-disc limit. Similar to our study, they did not rely on a typical MRI activation temperature but calculated an adaptive $\alpha$ based on a layered accretion model \citep[e.g.,][]{Gammie1996}, using a simple description of the ionisation fraction and accounting for the effect of Ohmic diffusivity. The most relevant differences between their setup and ours include the inner boundary being located much further away from the star (0.52~AU), their description of an initially constant mass to magnetic flux ratio and the evolution of the magnetic field in the ideal MHD limit (but without a stellar component), and their discs being in an earlier, embedded phase with infall. The bursts occurring in their simulations are not due to the dynamics around the DZIE, but rather due to the accumulation of mass as a consequence of a non-constant $\alpha$ within the dead zone. They also do not resolve the vertical disc structure and may therefore miss the development of Solberg--H{\o}iland-type instabilities that can extend the burst regions further. Despite the significant differences, the shapes of the individual bursts shown, for instance, in panel b of Fig. 7 of \citet{Vorobyov2020b}, are remarkably similar to the morphology of the RNM analysed in our work. The sharp increase in the accretion rate of the main flare is followed by a slower decline before abruptly dropping back to low values (even showing signs of decretion). Several weaker reflares occur before the accretion rate decreases below pre-burst levels. The same burst morphology results from our model \texttt{STAR2}, the accretion rate of which is shown in panel ($\mathrm{a_3}$) of Fig. \ref{fig:radius_time}. \citet{Vorobyov2020b} and \citet{Kadam2022} also tested less magnetised discs and found that the MRI active burst region generally does not expand as far as in cases with higher magnetisation, which is consistent with our findings. \par
Considering the structure of the inner disc inferred from MRI activity, one of the most recent models was presented by \citet{Jankovic2021} and \citet{Jankovic2022}, which extends the study by \citet{Mohanty2018}. They implemented the same MRI activation criteria as we do in this work, including the same sources of non-thermal ionisation. They further based their ionisation framework on the description by \citet{Desch2015}, the model from which we directly used the tabulated diffusivities. The most crucial difference between their setup and ours is the description of the magnetic field strength. While they chose the magnetic field strength at each radius such that $\bar\alpha$ (Eq. \ref{eq:alphabar}) is maximised, we a priori prescribed the magnetic field configuration consisting of a stellar and (fossil) disc component, and determined the resulting local MRI activity. Furthermore, \citet{Jankovic2021} calculated steady-state solutions without time-dependent dynamic evolution. However, as shown in \citetalias{Cecil2024b}, multiple solutions for the steady-state structure in terms of density and temperature can exist, forming the basis for the limit-cycle that describes the periodic MRI-triggered outbursts (described by S-curves of thermal equilibria, as shown in e.g. Fig. 11 of \citetalias{Cecil2026}). Since \citet{Jankovic2021} chose the solution that maximises the MRI activity, they always arrive at the solution corresponding to the outburst state in our models, where the MRI active region has reached its largest extent \footnote{More precisely, they found the mid-burst state of the (R)WM as described in our work.} \par
\cite{Jankovic2021} described the $\alpha$ parameter in the MRI active regime as $\alpha_\mathrm{MRI}=1/(3\beta)$, which follows from a relation derived from shearing box simulations by \citet{Bai2011}. Since this relation might be influenced by box sizes and boundary conditions \citep{Ross2016} and does not necessarily hold in global, stratified simulations \citep{Flock2013}, we chose to keep $\alpha_\mathrm{MRI}$ constant, following our previous work. \cite{Jankovic2021} found a radially very smooth transition between the MRI active and inactive regions, leading to a shallow increase of the surface density from 0.1~AU to the DZIE at $\sim\!\!0.7$~AU. Consequently, the density and pressure bump at the DZIE is less pronounced in their models compared to ours, where the transition manifests sharply (as especially apparent in the upper row of Fig. \ref{fig:sigma_profiles}).

\subsection{Long-term evolution} \label{sec:long_term}
The high computational load of our simulations only allowed us to follow the evolution of the inner disc for at most a few thousand years. The main reasons for these restrictions are the small radius of the inner boundary, the logarithmically spaced grid, and the multidimensional interpolations needed to extract opacities and diffusivities from lookup tables. Therefore, the long-term evolution of the disc over Myr timescales could not be captured. For this reason, we cannot make reliable statements about the exact number and frequency of outbursts that may occur over the disc's lifetime. Panel c of Fig. \ref{fig:radius_time} indicates that the quiescent phases between the bursts in the \texttt{STAR1} and \texttt{STAR2} models become longer with consecutive outbursts. While this is, in principle, expected in periodic instability models \citep[e.g.,][]{Vorobyov2006, Zhu2010b, Bae2013, Kadam2020}, it is also influenced by the choice of initial conditions in our case. \par
Panel c of Fig. \ref{fig:radius_time} also shows that, while the burst shapes and maximum accretion rates remain approximately unchanged, the accretion rates in each quiescent phase are lower than in the previous one. We expect this trend to continue until the quiescent accretion rate is set in the vicinity of $\dot{M}_\mathrm{init}$. The bursting mechanism should then continue (at a lower frequency) until the inner disc cannot be replenished to a state that meets the MRI activation criteria near the DZIE. To determine this cut-off point for the bursting mechanism, as well as the bursting frequency and long-term consequences for the inner planet-forming regions, it will be useful to combine our models with computationally more efficient methods. For instance, the 1D, fully implicit code \texttt{TAPIR} \citep{Ragossnig2020, Steiner2021} could be used to simulate the whole disc over its entire lifetime. Our 2D or 3D \texttt{PLUTO} models can then be implemented as zoom-in simulations when the conditions for an MRI-triggered outburst are met, with \texttt{TAPIR} providing the boundary conditions. Extracted metrics from the burst phase would then be fed back to \texttt{TAPIR}, which proceeds to model the quiescent phase until a new burst is triggered. With this method, not only can the MRI-triggered burst mechanism be resolved in great detail, but the long-term evolution of the entire disc can be captured as well.

\subsection{Model limitations}
Although our models include descriptions that significantly expand and improve upon previously published work, certain compromises had to be made to retain computational feasibility, model comparability, and clarity of the interplay among many physical processes. In this section, we briefly describe the main limitations of our models regarding magnetic field effects and dust treatment.

\subsubsection{Additional effects of magnetic fields}

The inner boundary of a PPD is commonly identified with the magnetic truncation radius, $r_\mathrm{t}$ \citep{Bessolaz2008, Hartmann2016}, defined as the location where the magnetic stress exerted by the stellar magnetosphere exceeds the ram pressure of the accreting disc material.
In the case of episodic accretion, $r_\mathrm{t}$ is strongly time-dependent; a substantial increase in the mass accretion rate drives $r_\mathrm{t}$ towards the stellar surface \citep{Steiner2021}.
This compresses the stellar magnetosphere, leading to the innermost disc becoming even hotter, which can, in turn, modify the bursting behaviour.
Additionally, a weaker stellar magnetic field moves $r_\mathrm{t}$ closer to the star, resulting in similar effects \citep{Steiner2021}. \par
The shift in $r_\mathrm{t}$ during accretion events can increase the temperature in the innermost regions further. In more massive discs than those considered in our study, this can lead to the emergence of classical thermal instability due to hydrogen ionisation during an outburst, triggering an instability cascade similar to that described by \citet{Kadam2020}. While the focus of our work is on the instability behaviour of the DZIE in relatively low-mass discs, without the contribution of classical thermal instability, a possible future prospect is to extend our models to even smaller radii and more massive discs. This would also allow for a better comparison of the accretion behaviour with observed light curves of outbursting FU Ori-type objects, which our current models are not designed for. \par

In addition to its dependence on the location of the inner disc rim, the disc structure is modified by magnetic torques exerted by the stellar magnetic field \citep[][]{Steiner2021}.
Inside the corotation radius, $r_\mathrm{cor}$, the disc gas is braked and accretes more rapidly onto the star, leading to a reduction in $\Sigma$.
Just outside $r_\mathrm{cor}$, the gas can accumulate, resulting in an increase in $\Sigma$ due to its acceleration \citep[the so-called slingshot effect; e.g.,][]{Steiner2021, Li2022}.
In addition, angular momentum exchange between the star and the innermost disc couples the stellar spin evolution to the disc evolution \citep{Gehrig2022, Gehrig2023, Gehrig2023MESA}.

Throughout this work, angular momentum is transported exclusively by turbulent viscosity.
However, magnetically driven outflows that couple to the large-scale magnetic field can also remove mass and angular momentum from the disc.
This process exerts an additional torque mediated by magnetic field lines anchored in the disc, thereby driving radial mass transport \citep{Bai2013,Lesur2023,Roberts2026}.
Moreover, the $\Sigma$ profile in the inner disc ($r \lesssim 5$~AU) may be altered by strong winds launched from the inner regions \citep[e.g.,][]{Cui2021}.
The occurrence and magnitude of magnetically driven outflows strongly depend on the topology and strength of the large-scale magnetic field \citep{Bai2013, Cui2021, Roberts2026}.

In this work, the disc magnetic field was assumed to be constant. 
Magnetic advection depends either on the turbulent viscosity present at a given radius or on the angular momentum removed by winds.
Magnetic diffusion, in turn, is strongly governed by the density and ionisation structure of the disc \citep[][]{Guilet2014, Kadam2025b, Roberts2026}.
While field-strength estimates based on the stationary profiles of \cite{Steiner2025} provide a reasonable approximation during the quiescent phases, the disc structure evolves rapidly during a burst.
Consequently, the large-scale magnetic field is also expected to vary, leading to changes in both its strength and topology.
Our results demonstrate that the disc magnetic field strength is a key parameter in determining the burst mode.
Therefore, incorporating a dynamically evolving disc magnetic field during a burst may give rise to an even broader range of bursting behaviours.\par
In our models, we assumed that the MRI is saturated when the two conditions in Eqs. \ref{eq:MRIcrit1} and \ref{eq:MRIcrit2} are met. But maximal MRI efficiency also requires that the fastest-growing MRI modes be allowed to develop in the active regions \citep[as considered, e.g., by][]{Delage2022}. Technically, this criterion is already included in the description of $\beta_\mathrm{min}$ given in Eq. \ref{eq:MRIcrit2}. However, the MRI active layers in our models can become thin (panel e of Fig. \ref{fig:structure}), approaching vertical extents comparable to one scale height. More detailed non-ideal MHD studies are needed to investigate whether the MRI can indeed operate efficiently in these thin layers.

\subsubsection{Dust and thermionic emission}
In our models, dust is considered only indirectly via the variable dust-to-gas mass ratio when determining the opacity of the disc material. We do not consider the growth, fragmentation, or distribution of dust grains. We refer to \citet{Ziampras2026} for a recent study of dust evolution in the framework of our MRI-triggered outburst mechanism and to \citet{Vorobyov2022} for an investigation of the response of dust at larger radii to strong accretion outbursts.\par
The lookup table of diffusivities we used for our models, which is based on the thermochemical model of \citet{Desch2015}, has been created for 1~$\mu$m-sized dust grains and a constant dust-to-gas mass ratio of $10^{-3}$ (which is the maximum dust-to-gas mass ratio, $f_0$, in our models). \citet{Desch2015} and, more recently, \citet{Williams2025} showed that the dust-to-gas ratio can have a strong effect on the ionisation state, at least at temperatures at which purely gas-phase reactions are not yet dominant. Additionally, \citet{Williams2025} found that different grain size distributions, as well as minimum and maximum grain sizes, can influence the resulting resistivities of non-ideal MHD effects. This is mainly due to their effect on the total surface area of dust grains, from which thermionic emission originates. While including these additional dimensions to determine MRI activity during the inner disc's highly variable evolution goes beyond the scope of our current study, future investigations can benefit greatly from considering these effects. \\

\section{Conclusion} \label{sec:conclusion}
With the models analysed in this work, we expanded on our previous investigations of the evolution of the inner regions of PPDs by implementing MRI activation criteria based on the non-ideal MHD effects of ambipolar and Ohmic diffusion. This allowed us to tie the MRI activity to prescribed magnetic field strength profiles. We showed the impact on the inner disc's structure and on the emergence, diverse manifestations, and consequences of MRI-triggered inner disc instabilities that result in episodic accretion bursts. In the following, we list the main results.

\begin{itemize}
    \item The detailed tracking of MRI activity reveals a new bursting mode. While the wide burst mode, previously observed in models describing the MRI activity via a simple activation temperature threshold, can still be reproduced by considering relatively strong stellar and disc magnetic fields, the narrow burst mode manifests only in weakly magnetised discs. 
    \item In the narrow mode, the MRI active region is restricted to small radii ($\lesssim 0.3$~AU in our models). While the pressure maximum at the DZIE is destroyed and recreated by the inner disc instability in the wide mode, it oscillates in the narrow mode due to more frequently repeating burst cycles and does not remain static during quiescence.
    \item Both narrow and wide modes can be further separated into reflaring and non-reflaring modes. Reflares are only occurring if the magnetic field near the inner disc edge is sufficiently strong (e.g. through a strong stellar magnetic field).
    \item The wide burst mode manifests if the magnetisation of the disc allows for the burst region to expand far enough to render the region around the ionisation front unstable to a Solberg-H{\o}iland-type instability. The hydrodynamic instability can then drive the heating front further outwards. This creates a distinct dichotomy of narrow and wide bursts.
    \item In contrast to the wide mode analysed in previous work, the narrow mode displays a sharp peak in the accretion rate, followed by an initially slow decline and, finally, an abrupt drop back to quiescent values. 
    \item The shape of the MRI active region in quiescence is significantly altered by the detailed activation criteria, compared to a simple $T_\mathrm{MRI}$ threshold. It is characterised by MRI active bands wedged between an Ohmic dead zone around the midplane and a dead region quenched by ambipolar diffusion in the upper layers. The bands converge toward the midplane near the star, where irradiation heating becomes dominant. During outbursts, the MRI active region closely resembles previously found solutions for steady-state structures that maximise MRI activity.
    \item In our 3D model, the density spike at the ionisation front in the narrow mode quickly becomes unstable to the RWI. This produces multiple vortices at a distance of $\sim\!\!0.3$~AU, which converge into a single vortex that diffuses over a timescale of around 150 local orbits.
\end{itemize}
The careful consideration of magnetic fields has repeatedly led to fundamental changes in our understanding of many astrophysical processes. The evolution and instability of the inner PPD are no exception. Our study indicates that the regions of a PPD, where the terrestrial planets of the Solar System and the majority of currently detected exoplanets are found, are highly dynamically variable and strongly depend on the presence and strength of magnetic fields. With the rapid advancement of numerical techniques and efficient codes, future studies will be able to couple these complex, multi-dimensional processes in the very inner regions with the disc's long-term evolution, enabling us to paint a more complete picture of the conditions for the formation and migration of planets close to their host star.

\begin{acknowledgements}
This research was supported by Deutsche Forschungsgemeinschaft (DFG, German Research Foundation) under grant no. 517644750. The authors are very grateful to Neal Turner and Steven Desch for providing the diffusivity tables.
Views and opinions expressed are those of the authors only.
\end{acknowledgements}
% ------------------------------------------------------------------
% FOOTER
% ------------------------------------------------------------------
\bibliographystyle{resources/aa}

\bibliography{literature/SETI_III}

@ARTICLE{Gehrig2023MESA,
       author = {{Gehrig}, L. and {Steindl}, T. and {Vorobyov}, E.~I. and {Guadarrama}, R. and {Zwintz}, K.},
        title = "{The influence of metallicity on a combined stellar and disk evolution}",
      journal = {A\&A},
         year = 2023,
        month = jan,
       volume = {669},
          eid = {A84},
        pages = {A84},
          doi = {10.1051/0004-6361/202244408},
archivePrefix = {arXiv},
       eprint = {2211.05331},
 primaryClass = {astro-ph.SR},
       adsurl = {https://ui.adsabs.harvard.edu/abs/2023A&A...669A..84G}
}

@ARTICLE{Gehrig2023,
       author = {{Gehrig}, L. and {Gaidos}, E. and {G{\"u}del}, M.},
        title = "{The post-disk (or primordial) spin distribution of M dwarf stars}",
      journal = {A\&A},
         year = 2023,
        month = jul,
       volume = {675},
          eid = {A179},
        pages = {A179},
          doi = {10.1051/0004-6361/202243521},
archivePrefix = {arXiv},
       eprint = {2306.02657},
 primaryClass = {astro-ph.SR},
       adsurl = {https://ui.adsabs.harvard.edu/abs/2023A&A...675A.179G}
}

@ARTICLE{Gehrig2022,
       author = {{Gehrig}, L. and {Steiner}, D. and {Vorobyov}, E.~I. and {G{\"u}del}, M.},
        title = "{Time-dependent, long-term hydrodynamic simulations of the inner protoplanetary disk. II. The importance of stellar rotation}",
      journal = {A\&A},
         year = 2022,
        month = nov,
       volume = {667},
          eid = {A46},
        pages = {A46},
          doi = {10.1051/0004-6361/202243549},
archivePrefix = {arXiv},
       eprint = {2208.08852},
 primaryClass = {astro-ph.SR},
       adsurl = {https://ui.adsabs.harvard.edu/abs/2022A&A...667A..46G}
}

@ARTICLE{Kadam2025b,
       author = {{Kadam}, Kundan and {Vorobyov}, Eduard and {Woitke}, Peter and {Basu}, Shantanu and {van Terwisga}, Sierk},
        title = "{Magnetic disk winds in protoplanetary disks: Description of the model and impact on global disk evolution}",
      journal = {A\&A},
         year = 2025,
        month = mar,
       volume = {695},
          eid = {A167},
        pages = {A167},
          doi = {10.1051/0004-6361/202450236},
archivePrefix = {arXiv},
       eprint = {2502.00161},
 primaryClass = {astro-ph.EP},
       adsurl = {https://ui.adsabs.harvard.edu/abs/2025A&A...695A.167K}
}

@ARTICLE{Cui2021,
       author = {{Cui}, Can and {Bai}, Xue-Ning},
        title = "{Global three-dimensional simulations of outer protoplanetary discs with ambipolar diffusion}",
      journal = {\mnras},
         year = 2021,
        month = oct,
       volume = {507},
       number = {1},
        pages = {1106-1126},
          doi = {10.1093/mnras/stab2220},
archivePrefix = {arXiv},
       eprint = {2106.10167},
 primaryClass = {astro-ph.EP},
       adsurl = {https://ui.adsabs.harvard.edu/abs/2021MNRAS.507.1106C}
}

@article{Guilet2014,

archivePrefix = {arXiv},
arxivId = {1403.3732},
author = {Guilet, Jerome and Ogilvie, Gordon I.},
doi = {10.1093/mnras/stu532},
eprint = {1403.3732},
issn = {0035-8711},
journal = {MNRAS},
month = {mar},
pages = {852--868},
title = {{Global evolution of the magnetic field in a thin disc and its consequences for protoplanetary systems}},
url = {http://arxiv.org/abs/1403.3732 http://dx.doi.org/10.1093/mnras/stu532},
volume = {441},
year = {2014}
}

@ARTICLE{Bessolaz2008,
       author = {{Bessolaz}, N. and {Zanni}, C. and {Ferreira}, J. and {Keppens}, R. and
         {Bouvier}, J.},
        title = "{Accretion funnels onto weakly magnetized young stars}",
      journal = {\aap},
         year = 2008,
        month = jan,
       volume = {478},
       number = {1},
        pages = {155-162},
          doi = {10.1051/0004-6361:20078328},
archivePrefix = {arXiv},
       eprint = {0712.2921},
 primaryClass = {astro-ph},
       adsurl = {https://ui.adsabs.harvard.edu/abs/2008A&A...478..155B}
}

@article{Ziampras2026,
  
   author = {Alexandros Ziampras and Tilman Birnstiel and Nicolas Kaufmann and Michael Cecil and Thomas Pfeil and Alexandros Ziampras and Tilman Birnstiel and Nicolas Kaufmann and Michael Cecil and Thomas Pfeil and Ziampras and Alexandros and Birnstiel and Tilman and Kaufmann and Nicolas and Cecil and Michael and Pfeil and Thomas},
   journal = {},
   month = {2},
   pages = {arXiv:2602.20283, accepted by A\&A},
   title = {Planet formation at the inner edge of the dead zone -- I: the interplay between accretion outbursts and dust growth},
   url = {https://ui.adsabs.harvard.edu/abs/2026arXiv260220283Z/abstract},
   year = {2026}
}

@article{Cecil2026,
   author = {M. Cecil and M. Flock and M. G. Malygin and R. Kuiper and P. Sudarshan and A. Ziampras and V. G. Elbakyan},
   doi = {10.1051/0004-6361},
   issn = {0004-6361},
   journal = {A\&A},
   pages = {A296},
   title = {The role of detailed gas and dust opacities in shaping the evolution of the inner disc edge subject to episodic accretion},
   volume = {707},
   url = {https://ui.adsabs.harvard.edu/abs/2026A&A...707A.296C/abstract},
   year = {2026}
}

@article{Vanderburg2016,
   author = {Andrew Vanderburg and Peter Plavchan and John Asher Johnson and David R. Ciardi and Jonathan Swift and Stephen R. Kane},
   doi = {10.1093/mnras/stw863},
   issn = {13652966},
   issue = {4},
   journal = {MNRAS},
   month = {7},
   pages = {3565-3573},
   publisher = {Oxford University Press},
   title = {Radial velocity planet detection biases at the stellar rotational period},
   volume = {459},
   url = {https://ui.adsabs.harvard.edu/abs/2016MNRAS.459.3565V/abstract},
   year = {2016}
}

@article{Hara2023,
   author = {Nathan C. Hara and Eric B. Ford},
   doi = {10.1146/annurev-statistics-033021-012225},
   issn = {2326831X},
   issue = {1},
   journal = {Annu. Rev. Stat. Appl.},
   month = {3},
   pages = {623-649},
   publisher = {Annual Reviews Inc.},
   title = {Statistical Methods for Exoplanet Detection with Radial Velocities},
   volume = {10},
   url = {https://ui.adsabs.harvard.edu/abs/2023AnRSA..10..623H/abstract},
   year = {2023}
}

@article{Ananyeva2023,
   author = {Vladislava Ananyeva and Anastasiia Ivanova and Inna Shashkova and Oleg Yakovlev and Alexander Tavrov and Oleg Korablev and Jean-Loup Bertaux and Vladislava Ananyeva and Anastasiia Ivanova and Inna Shashkova and Oleg Yakovlev and Alexander Tavrov and Oleg Korablev and Jean-Loup Bertaux and Ananyeva and Vladislava and Ivanova and Anastasiia and Shashkova and Inna and Yakovlev and Oleg and Tavrov and Alexander and Korablev and Oleg and Bertaux and Jean-Loup},
   doi = {10.3390/atmos14020353},
   issn = {20734433},
   issue = {2},
   journal = {Atmos},
   month = {2},
   pages = {353},
   publisher = {MDPI},
   title = {Exoplanets Catalogue Analysis: The Distribution of Exoplanets at FGK Stars by Mass and Orbital Period Accounting for the Observational Selection in the Radial Velocity Method},
   volume = {14},
   url = {https://ui.adsabs.harvard.edu/abs/2023Atmos..14..353A/abstract},
   year = {2023}
}

@article{Roberts2026,
   author = {Matthew J. O. Roberts and Henrik N. Latter and Geoffroy Lesur},
   journal = {MNRAS},
   month = {2},
   pages = {1-24},
   title = {Global magnetohydrodynamic simulations of the inner regions of protoplanetary discs. II. Vertical-net-flux regime},
   volume = {000},
   url = {https://arxiv.org/pdf/2602.11818},
   year = {2026}
}

@article{Ragossnig2020,
   author = {Florian Ragossnig and Ernst A. Dorfi and Bernhard Ratschiner and Lukas Gehrig and Daniel Steiner and Alexander Stökl and Colin P. Johnstone},
   doi = {10.1016/j.cpc.2020.107437},
   issn = {00104655},
   journal = {Comp. Phys. Commun.},
   month = {11},
   pages = {107437},
   publisher = {Elsevier B.V.},
   title = {1+1D implicit disk computations},
   volume = {256},
   url = {http://arxiv.org/abs/2006.12939 http://dx.doi.org/10.1016/j.cpc.2020.107437},
   year = {2020}
}

@article{Ross2016,
   author = {Johnathan Ross and Henrik N. Latter and Jerome Guilet},
   doi = {10.1093/mnras/stv2286},
   issn = {13652966},
   issue = {1},
   journal = {MNRAS},
   month = {1},
   pages = {526-539},
   publisher = {Oxford University Press},
   title = {The stress-pressure relationship in simulations of MRI-induced turbulence},
   volume = {455},
   url = {https://ui.adsabs.harvard.edu/abs/2016MNRAS.455..526R/abstract},
   year = {2016}
}

@article{Hu2026,
   author = {Xiao Hu and Jonathan C. Tan},
   doi = {10.3847/1538-4357/AE4020},
   issn = {0004-637X},
   issue = {2},
   journal = {ApJ},
   month = {3},
   pages = {220},
   title = {Inside-out Planet Formation. VIII. Onset of Planet Formation and the Transition Disk Phase},
   volume = {999},
   url = {https://ui.adsabs.harvard.edu/abs/2026ApJ...999..220H/abstract},
   year = {2026}
}

@article{Machida2019,
   author = {Masahiro N. Machida and Shantanu Basu},
   doi = {10.3847/1538-4357/ab18a7},
   issn = {0004-637X},
   issue = {2},
   journal = {ApJ},
   month = {5},
   pages = {149},
   publisher = {American Astronomical Society},
   title = {The First Two Thousand Years of Star Formation},
   volume = {876},
   url = {https://ui.adsabs.harvard.edu/abs/2019ApJ...876..149M/abstract},
   year = {2019}
}

@article{Vaytet2018,
   author = {N. Vaytet and B. Commerçon and J. Masson and M. González and G. Chabrier},
   doi = {10.1051/0004-6361/201732075},
   issn = {14320746},
   journal = {A\&A},
   month = {7},
   pages = {A5},
   publisher = {EDP Sciences},
   title = {Protostellar birth with ambipolar and ohmic diffusion},
   volume = {615},
   url = {https://ui.adsabs.harvard.edu/abs/2018A&A...615A...5V/abstract},
   year = {2018}
}

@article{Weiss2021,
   author = {Benjamin P. Weiss and Xue-Ning Bai and Roger R. Fu},
   doi = {10.1126/sciadv.aba5967},
   issue = {1},
   journal = {SciA},
   month = {3},
   pages = {eaba5967},
   publisher = {American Association for the Advancement of Science},
   title = {History of the Solar Nebula from Meteorite Paleomagnetism},
   volume = {7},
   url = {http://arxiv.org/abs/2103.02011 http://dx.doi.org/10.1126/sciadv.aba5967},
   year = {2021}
}

@article{Fu2023,
   author = {Roger R. Fu and Sarah C. Steele and Jacob B. Simon and Richard Teague and Joan Najita and David Rea and Roger R. Fu and Sarah C. Steele and Jacob B. Simon and Richard Teague and Joan Najita and David Rea and Fu and Roger R. and Steele and Sarah C. and Simon and Jacob B. and Teague and Richard and Najita and Joan and Rea and David},
   doi = {10.3847/PSJ/ace716},
   issn = {26323338},
   issue = {8},
   journal = {PSJ},
   month = {8},
   pages = {151},
   publisher = {Institute of Physics},
   title = {Implications for Chondrule Formation Regions and Solar Nebula Magnetism from Statistical Reanalysis of Chondrule Paleomagnetism},
   volume = {4},
   url = {https://ui.adsabs.harvard.edu/abs/2023PSJ.....4..151F/abstract},
   year = {2023}
}

@article{Ahmad2025,
   author = {Adnan Ali Ahmad and Matthias González and Patrick Hennebelle and Ugo Lebreuilly and Benoît Commerçon},
   doi = {10.1051/0004-6361/202553663},
   journal = {A\&A},
   month = {3},
   pages = {A238},
   publisher = {EDP Sciences},
   title = {Birth of magnetized low-mass protostars and circumstellar disks},
   volume = {696},
   url = {http://arxiv.org/abs/2503.08637 http://dx.doi.org/10.1051/0004-6361/202553663},
   year = {2025}
}

@article{Masson2015,
   author = {Jacques Masson and Gilles Chabrier and Patrick Hennebelle and Neil Vaytet and Benoit Commerçon},
   doi = {10.1051/0004-6361/201526371},
   journal = {A\&A},
   month = {10},
   pages = {A32},
   publisher = {EDP Sciences},
   title = {Ambipolar diffusion in low-mass star formation. I. General comparison with the ideal MHD case},
   volume = {587},
   url = {http://arxiv.org/abs/1509.05630 http://dx.doi.org/10.1051/0004-6361/201526371},
   year = {2015}
}

@article{Woo2023,
   author = {J. M.Y. Woo and A. Morbidelli and S. L. Grimm and J. Stadel and R. Brasser},
   doi = {10.1016/j.icarus.2023.115497},
   issn = {10902643},
   journal = {Icarus},
   month = {5},
   pages = {115497},
   publisher = {Academic Press Inc.},
   title = {Terrestrial planet formation from a ring},
   volume = {396},
   url = {http://arxiv.org/abs/2302.14100 http://dx.doi.org/10.1016/j.icarus.2023.115497},
   year = {2023}
}

@article{Nesvorn2025,
   author = {David Nesvorný and Alessandro Morbidelli and William F. Bottke and Rogerio Deienno and Max Goldberg and David Nesvorný and Alessandro Morbidelli and William F. Bottke and Rogerio Deienno and Max Goldberg and Nesvorný and David and Morbidelli and Alessandro and Bottke and William F. and Deienno and Rogerio and Goldberg and Max},
   doi = {10.3847/1538-3881/adf20a},
   issn = {0004-6256},
   issue = {3},
   journal = {AJ},
   month = {9},
   pages = {180},
   publisher = {American Astronomical Society},
   title = {Terrestrial Planet Formation from Two Source Reservoirs},
   volume = {170},
   url = {https://ui.adsabs.harvard.edu/abs/2025AJ....170..180N/abstract},
   year = {2025}
}

@article{Paardekooper2022,
   author = {Sijme-Jan Paardekooper and Ruobing Dong and Paul Duffell and Jeffrey Fung and Frederic S. Masset and Gordon Ogilvie and Hidekazu Tanaka},
   journal = {ASPC},
   month = {3},
   pages = {685},
   title = {Planet-Disk Interactions},
   volume = {534},
   url = {http://arxiv.org/abs/2203.09595},
   year = {2022}
}

@article{Khaibrakhmanov2024,
   author = {S. A. Khaibrakhmanov},
   doi = {10.17184/eac.8855},
   issn = {14763540},
   issue = {2},
   journal = {AApTr},
   month = {7},
   pages = {139-162},
   publisher = {Cambridge Scientific Publishers},
   title = {Magnetic fields of protoplanetary disks},
   volume = {34},
   url = {https://ui.adsabs.harvard.edu/abs/2024A&AT...34..139K/abstract},
   year = {2024}
}

@article{Dudorov2022,
   author = {A. E. Dudorov and S. N. Zamozdra},
   doi = {10.1134/S1063772922040035},
   issn = {15626881},
   issue = {3},
   journal = {ARep},
   month = {3},
   pages = {200-220},
   publisher = {Pleiades journals},
   title = {Effect of Dust Evaporation on the Fossil Magnetic Field of Young Stars and Their Accretion Disks},
   volume = {66},
   url = {https://ui.adsabs.harvard.edu/abs/2022ARep...66..200D/abstract},
   year = {2022}
}

@article{Teague2025,
   author = {Richard Teague and Boy Lankhaar and Sean M. Andrews and Chunhua Qi and Roger R. Fu and David J. Wilner and John B. Biersteker and Joan R. Najita},
   doi = {10.3847/2041-8213/adff4d},
   isbn = {2022.1.00840},
   issn = {2041-8205},
   issue = {1},
   journal = {ApJL},
   month = {9},
   pages = {L6},
   publisher = {American Astronomical Society},
   title = {A Radially Resolved Magnetic Field Threading the Disk of TW Hya},
   volume = {991},
   url = {https://ui.adsabs.harvard.edu/abs/2025ApJ...991L...6T/abstract},
   year = {2025}
}

@article{Masset2016,
   author = {F. S. Masset and P. Benítez-Llambay},
   doi = {10.3847/0004-637x/817/1/19},
   issn = {0004-637X},
   issue = {1},
   journal = {ApJ},
   month = {1},
   pages = {19},
   publisher = {IOP Publishing},
   title = {HORSESHOE DRAG IN THREE-DIMENSIONAL GLOBALLY ISOTHERMAL DISKS},
   volume = {817},
   url = {https://iopscience.iop.org/article/10.3847/0004-637X/817/1/19 https://iopscience.iop.org/article/10.3847/0004-637X/817/1/19/meta},
   year = {2016}
}

@article{Fu2021,
   author = {Roger R. Fu and Michael W. R. Volk and Dario Bilardello and Guy Libourel and Geoffroy R. J. Lesur and Oren Ben Dor and Roger R. Fu and Michael W. R. Volk and Dario Bilardello and Guy Libourel and Geoffroy R. J. Lesur and Oren Ben Dor and Fu and Roger R. and Volk and Michael W. R. and Bilardello and Dario and Libourel and Guy and Lesur and Geoffroy R. J. and Ben Dor and Oren},
   doi = {10.1029/2021av000486},
   issn = {2576-604X},
   issue = {3},
   journal = {AGUA},
   month = {9},
   pages = {e00486},
   publisher = {American Geophysical Union (AGU)},
   title = {The Fine-Scale Magnetic History of the Allende Meteorite: Implications for the Structure of the Solar Nebula},
   volume = {2},
   url = {https://ui.adsabs.harvard.edu/abs/2021AGUA....200486F/abstract},
   year = {2021}
}

@article{Fu2014,
   author = {Roger R. Fu and Benjamin P. Weiss and Eduardo A. Lima and Richard J. Harrison and Xue-Ning Bai and Steven J. Desch and Denton S. Ebel and Clément Suavet and Huapei Wang and David Glenn and David Le Sage and Takeshi Kasama and Ronald L. Walsworth and Aaron T. Kuan and Roger R. Fu and Benjamin P. Weiss and Eduardo A. Lima and Richard J. Harrison and Xue-Ning Bai and Steven J. Desch and Denton S. Ebel and Clément Suavet and Huapei Wang and David Glenn and David Le Sage and Takeshi Kasama and Ronald L. Walsworth and Aaron T. Kuan and Fu and Roger R. and Weiss and Benjamin P. and Lima and Eduardo A. and Harrison and Richard J. and Bai and Xue-Ning and Desch and Steven J. and Ebel and Denton S. and Suavet and Clément and Wang and Huapei and Glenn and David and Le Sage and David and Kasama and Takeshi and Walsworth and Ronald L. and Kuan and Aaron T.},
   doi = {10.1126/science.1258022},
   issn = {10959203},
   issue = {6213},
   journal = {Sci},
   month = {11},
   pages = {1089-1092},
   publisher = {American Association for the Advancement of Science},
   title = {Solar nebula magnetic fields recorded in the Semarkona meteorite},
   volume = {346},
   url = {https://www.science.org/doi/pdf/10.1126/science.1258022?download=true},
   year = {2014}
}

@article{Mansbach2024,
   author = {Elias N. Mansbach and Benjamin P. Weiss and Eduardo A. Lima and Michael Sowell and Joseph L. Kirschvink and Roger R. Fu and Saverio Cambioni and Xue-Ning Bai and Jodie B. Ream and Chisato Anai and Atsuko Kobayashi and Hironori Hidaka and Elias N. Mansbach and Benjamin P. Weiss and Eduardo A. Lima and Michael Sowell and Joseph L. Kirschvink and Roger R. Fu and Saverio Cambioni and Xue-Ning Bai and Jodie B. Ream and Chisato Anai and Atsuko Kobayashi and Hironori Hidaka and Mansbach and Elias N. and Weiss and Benjamin P. and Lima and Eduardo A. and Sowell and Michael and Kirschvink and Joseph L. and Fu and Roger R. and Cambioni and Saverio and Bai and Xue-Ning and Ream and Jodie B. and Anai and Chisato and Kobayashi and Atsuko and Hidaka and Hironori},
   doi = {10.1029/2024AV001396},
   issn = {2576604X},
   issue = {6},
   journal = {AGUA},
   month = {12},
   pages = {2024AV001396},
   publisher = {John Wiley and Sons Inc},
   title = {Evidence for Magnetically-Driven Accretion in the Distal Solar System},
   volume = {5},
   url = {https://ui.adsabs.harvard.edu/abs/2024AGUA....501396M/abstract},
   year = {2024}
}

@article{Maurel2024,
   author = {Clara Maurel and Jérôme Gattacceca and Clara Maurel and Jérôme Gattacceca and Maurel and Clara and Gattacceca and Jérôme},
   doi = {10.1073/pnas.2312802121},
   issn = {10916490},
   issue = {12},
   journal = {PNAS},
   month = {3},
   pages = {e2312802121},
   pmid = {38437531},
   publisher = {National Academy of Sciences},
   title = {A 4,565-My-old record of the solar nebula field},
   volume = {121},
   url = {https://ui.adsabs.harvard.edu/abs/2024PNAS..12112802M/abstract},
   year = {2024}
}

@article{Roberts2025,
   author = {Matthew J.O. Roberts and Henrik N. Latter and Geoffroy Lesur},
   doi = {10.1093/mnras/staf1672},
   issn = {13652966},
   issue = {1},
   journal = {MNRAS},
   month = {11},
   pages = {1284-1303},
   publisher = {Oxford University Press},
   title = {Global magnetohydrodynamic simulations of the inner regions of protoplanetary discs. I. Zero-net flux regime},
   volume = {544},
   url = {http://arxiv.org/abs/2506.16945 http://dx.doi.org/10.1093/mnras/staf1672},
   year = {2025}
}

@article{Gdel2007,
   author = {M. Güdel and K. R. Briggs and K. Arzner and M. Audard and J. Bouvier and E. D. Feigelson and E. Franciosini and A. Glauser and N. Grosso and G. Micela and J. L. Monin and T. Montmerle and D. L. Padgett and F. Palla and I. Pillitteri and L. Rebull and L. Scelsi and B. Silva and S. L. Skinner and B. Stelzer and A. Telleschi},
   doi = {10.1051/0004-6361:20065724},
   issn = {00046361},
   issue = {2},
   journal = {A\&A},
   month = {6},
   pages = {353-377},
   title = {The XMM-Newton extended survey of the Taurus molecular cloud (XEST)},
   volume = {468},
   url = {https://ui.adsabs.harvard.edu/abs/2007A&A...468..353G/abstract},
   year = {2007}
}

@article{Igea1999,
   author = {J. Igea and A. E. Glassgold},
   doi = {10.1086/307302},
   issn = {0004-637X},
   issue = {2},
   journal = {ApJ},
   month = {6},
   pages = {848-858},
   publisher = {American Astronomical Society},
   title = {X-Ray Ionization of the Disks of Young Stellar Objects},
   volume = {518},
   url = {https://ui.adsabs.harvard.edu/abs/1999ApJ...518..848I/abstract},
   year = {1999}
}

@article{Bai2009,
   author = {Xue-Ning Bai and Jeremy Goodman},
   doi = {10.1088/0004-637X/701/1/737},
   issue = {1},
   journal = {ApJ},
   month = {7},
   pages = {737-755},
   publisher = {Institute of Physics Publishing},
   title = {Heat and Dust in Active Layers of Protostellar Disks},
   volume = {701},
   url = {http://arxiv.org/abs/0904.1240 http://dx.doi.org/10.1088/0004-637X/701/1/737},
   year = {2009}
}

@article{Dudorov2014,
   author = {A. E. Dudorov and S. A. Khaibrakhmanov},
   doi = {10.1007/s10509-014-1900-4},
   issue = {1},
   journal = {Ap\&SS},
   month = {3},
   pages = {103-121},
   publisher = {Springer Netherlands},
   title = {Fossil magnetic field of accretion disks of young stars},
   volume = {352},
   url = {http://arxiv.org/abs/1403.5513 http://dx.doi.org/10.1007/s10509-014-1900-4},
   year = {2014}
}

@article{Bai2011,
   author = {Xue Ning Bai and James M. Stone},
   doi = {10.1088/0004-637X/736/2/144},
   issn = {15384357},
   issue = {2},
   journal = {ApJ},
   month = {8},
   pages = {144},
   publisher = {Institute of Physics Publishing},
   title = {Effect of ambipolar diffusion on the nonlinear evolution of magnetorotational instability in weakly ionized disks},
   volume = {736},
   url = {https://ui.adsabs.harvard.edu/abs/2011ApJ...736..144B/abstract},
   year = {2011}
}

@article{Teed2021,
   author = {Robert J. Teed and Henrik N. Latter},
   doi = {10.1093/mnras/stab2311},
   issn = {13652966},
   issue = {4},
   journal = {MNRAS},
   month = {11},
   pages = {5523-5541},
   publisher = {Oxford University Press},
   title = {Axisymmetric simulations of the convective overstability in protoplanetary discs},
   volume = {507},
   url = {https://ui.adsabs.harvard.edu/abs/2021MNRAS.507.5523T/abstract},
   year = {2021}
}

@article{Latter2016,
   author = {Henrik N. Latter},
   doi = {10.1093/mnras/stv2449},
   issn = {13652966},
   issue = {3},
   journal = {MNRAS},
   month = {1},
   pages = {2608-2618},
   publisher = {Oxford University Press},
   title = {On the convective overstability in protoplanetary discs},
   volume = {455},
   url = {https://ui.adsabs.harvard.edu/abs/2016MNRAS.455.2608L/abstract},
   year = {2016}
}

@article{Klahr2014,
   author = {Hubert Klahr and Alexander Hubbard},
   doi = {10.1088/0004-637X/788/1/21},
   issn = {15384357},
   issue = {1},
   journal = {ApJ},
   month = {6},
   pages = {21},
   publisher = {Institute of Physics Publishing},
   title = {Convective overstability in radially stratified accretion disks under thermal relaxation},
   volume = {788},
   url = {https://ui.adsabs.harvard.edu/abs/2014ApJ...788...21K/abstract},
   year = {2014}
}

@article{Vorobyov2020b,
   author = {Eduard I. Vorobyov and Sergey Khaibrakhmanov and Shantanu Basu and Marc Audard},
   doi = {10.1051/0004-6361/202039081},
   issn = {14320746},
   journal = {A\&A},
   month = {12},
   pages = {A74},
   publisher = {EDP Sciences},
   title = {Accretion bursts in magnetized gas-dust protoplanetary disks},
   volume = {644},
   url = {https://ui.adsabs.harvard.edu/abs/2020A&A...644A..74V/abstract},
   year = {2020}
}

@article{Steiner2025,
   author = {D. Steiner and L. Gehrig and M. Güdel},
   doi = {10.1051/0004-6361/202554871},
   journal = {A\&A},
   month = {9},
   pages = {A163},
   publisher = {EDP Sciences},
   title = {Protoplanetary disks around magnetized young stars with large-scale magnetic fields I: Steady-state solutions},
   volume = {703},
   url = {http://arxiv.org/abs/2509.08393 http://dx.doi.org/10.1051/0004-6361/202554871},
   year = {2025}
}

@article{Cleeves2013,
   author = {L. Ilsedore Cleeves and Fred C. Adams and Edwin A. Bergin and Ruud Visser},
   doi = {10.1088/0004-637X/777/1/28},
   issn = {15384357},
   issue = {1},
   journal = {ApJ},
   month = {11},
   pages = {28},
   publisher = {Institute of Physics Publishing},
   title = {Radionuclide ionization in protoplanetary disks: Calculations of decay product radiative transfer},
   volume = {777},
   url = {https://ui.adsabs.harvard.edu/abs/2013ApJ...777...28C/abstract},
   year = {2013}
}

@article{Umebayashi2009,
   author = {Toyoharu Umebayashi and Takenori Nakano},
   doi = {10.1088/0004-637X/690/1/69},
   issn = {0004-637X},
   issue = {1},
   journal = {ApJ},
   pages = {69-81},
   publisher = {Institute of Physics Publishing},
   title = {Effects of Radionuclides on the Ionization State of Protoplanetary Disks and Dense Cloud Cores},
   volume = {690},
   url = {https://ui.adsabs.harvard.edu/abs/2009ApJ...690...69U/abstract},
   year = {2009}
}

@article{Delage2022,
   author = {Timmy N. Delage and Satoshi Okuzumi and Mario Flock and Paola Pinilla and Natalia Dzyurkevich},
   doi = {10.1051/0004-6361/202141689},
   issn = {14320746},
   journal = {A\&A},
   month = {2},
   pages = {A97},
   publisher = {EDP Sciences},
   title = {Steady-state accretion in magnetized protoplanetary disks},
   volume = {658},
   url = {https://ui.adsabs.harvard.edu/abs/2022A&A...658A..97D/abstract},
   year = {2022}
}

@article{Pea2025,
   author = {C. Contreras Peña and J. -E. Lee and G. Herczeg and D. Johnstone and P. Ábrahám and S. Antoniucci and M. Audard and M. Ashraf and G. Baek and A. Caratti o Garatti and A. Carvalho and L. Cieza and F. Cruz-Saénz de Miera and J. Eislöffel and D. Froebrich and T. Giannini and J. Green and A. Ghosh and Z. Guo and L. Hillenbrand and K. Hodapp and H. Jheonn and J. Jose and Y. -J. Kim and A. Kospál and H. -G. Lee and P. W. Lucas and T. Magakian and Z. Nagy and T. Naylor and J. P. Ninan and S. Peneva and Bo Reipurth and A. Scholz and E. Semkov and A. Sicilia-Aguilar and K. Singh and M. Siwak and B. Stecklum and Z. M. Szabó and V. Wolf and S. -Y. Yoon},
   doi = {10.5303/JKAS.2025.58.2.209},
   issue = {2},
   journal = {JKAS},
   month = {9},
   pages = {209-230},
   title = {The Outbursting YSOs Catalogue (OYCAT)},
   volume = {58},
   url = {http://arxiv.org/abs/2509.24876 http://dx.doi.org/10.5303/JKAS.2025.58.2.209},
   year = {2025}
}

@article{Elbakyan2025,
   author = {Vardan Elbakyan and Dennis Wehner and Rolf Kuiper and Sergei Nayakshin and Alessio Caratti o Garatti and Zhen Guo},
   doi = {10.1051/0004-6361/202555559},
   journal = {A\&A},
   month = {7},
   pages = {A91},
   publisher = {EDP Sciences},
   title = {Episodic accretion in high-mass star formation: An analysis of thermal instability for axially symmetric disks},
   volume = {701},
   url = {http://arxiv.org/abs/2507.20781 http://dx.doi.org/10.1051/0004-6361/202555559},
   year = {2025}
}

@article{Kadam2020,
   author = {Kundan Kadam and Eduard Vorobyov and Zsolt Regály and {\'A}gnes Kóspál and Péter Ábrahám},
   doi = {10.3847/1538-4357/ab8bd8},
   issn = {0004-637X},
   issue = {1},
   journal = {ApJ},
   month = {5},
   pages = {41},
   publisher = {American Astronomical Society},
   title = {Outbursts in Global Protoplanetary Disk Simulations},
   volume = {895},
   url = {https://ui.adsabs.harvard.edu/abs/2020ApJ...895...41K/abstract},
   year = {2020}
}

@article{Dullemond2018,
   author = {Cornelis P. Dullemond and Tilman Birnstiel and Jane Huang and Nicolás T. Kurtovic and Sean M. Andrews and Viviana V. Guzmán and Laura M. Pérez and Andrea Isella and Zhaohuan Zhu and Myriam Benisty and David J. Wilner and Xue-Ning Bai and John M. Carpenter and Shangjia Zhang and Luca Ricci},
   doi = {10.3847/2041-8213/aaf742},
   issn = {2041-8205},
   issue = {2},
   journal = {ApJL},
   month = {12},
   pages = {L46},
   publisher = {American Astronomical Society},
   title = {The Disk Substructures at High Angular Resolution Project (DSHARP). VI. Dust Trapping in Thin-ringed Protoplanetary Disks},
   volume = {869},
   url = {https://ui.adsabs.harvard.edu/abs/2018ApJ...869L..46D/abstract},
   year = {2018}
}

@article{Williams2025,
   author = {Morgan Williams and Subhanjoy Mohanty},
   doi = {10.1093/mnras/stae2510},
   issue = {2},
   journal = {MNRAS},
   month = {11},
   pages = {1518-1537},
   publisher = {Oxford University Press},
   title = {Ionization chemistry in the inner disc: a combined treatment of ionic and thermionic emission and arbitrary grain size distributions},
   volume = {536},
   url = {http://arxiv.org/abs/2411.14640 http://dx.doi.org/10.1093/mnras/stae2510},
   year = {2025}
}

@article{Iwasaki2024,
   author = {Kazunari Iwasaki and Kengo Tomida and Shinsuke Takasao and Satoshi Okuzumi and Takeru K. Suzuki},
   doi = {10.1093/pasj/psae036},
   issue = {4},
   journal = {PASJ},
   month = {6},
   pages = {616-652},
   publisher = {Oxford University Press},
   title = {Dynamics Near the Inner Dead-Zone Edges in a Proprotoplanetary Disk},
   volume = {76},
   url = {http://arxiv.org/abs/2401.03733 http://dx.doi.org/10.1093/pasj/psae036},
   year = {2024}
}

@article{Das2025,
   author = {Indrani Das and Eduard Vorobyov and Shantanu Basu},
   doi = {10.3847/1538-4357/adb8ee},
   issue = {2},
   journal = {ApJ},
   month = {2},
   pages = {163},
   publisher = {American Astronomical Society},
   title = {Accretion bursts in young intermediate-mass stars make planet formation challenging},
   volume = {983},
   url = {http://arxiv.org/abs/2502.17114 http://dx.doi.org/10.3847/1538-4357/adb8ee},
   year = {2025}
}

@article{Elbakyan2024,
   author = {Vardan G. Elbakyan and Sergei Nayakshin and Alessio Caratti o Garatti and Rolf Kuiper and Zhen Guo},
   doi = {10.1051/0004-6361/202451758},
   journal = {A\&A},
   month = {11},
   pages = {A256},
   publisher = {EDP Sciences},
   title = {The Role of Thermal Instability in Accretion Outbursts in High-Mass Stars},
   volume = {692},
   url = {http://arxiv.org/abs/2411.06949 http://dx.doi.org/10.1051/0004-6361/202451758},
   year = {2024}
}

@article{Yang2010,
   author = {Chao-Chin Yang and Kristen Menou},
   doi = {10.1111/j.1365-2966.2009.16047.x},
   issue = {4},
   journal = {MNRAS},
   month = {11},
   pages = {2436-2440},
   publisher = {Blackwell Publishing Ltd},
   title = {Rayleigh Adjustment of Narrow Barriers in Protoplanetary Discs},
   volume = {402},
   url = {http://arxiv.org/abs/0904.4266 http://dx.doi.org/10.1111/j.1365-2966.2009.16047.x},
   year = {2010}
}

@article{Chandrasekhar1961,
   author = {Subrahmanyan Chandrasekhar},
   journal = {Clarendon Press, Oxford},
   title = {Hydrodynamic and hydromagnetic stability},
   url = {https://ui.adsabs.harvard.edu/abs/1961hhs..book.....C/abstract},
   year = {1961}
}

@article{Woitke2016,
   author = {P. Woitke and M. Min and C. Pinte and W. F. Thi and I. Kamp and C. Rab and F. Anthonioz and S. Antonellini and C. Baldovin-Saavedra and A. Carmona and C. Dominik and O. Dionatos and J. Greaves and M. Güdel and J. D. Ilee and A. Liebhart and F. Ménard and L. Rigon and L. B.F.M. Waters and G. Aresu and R. Meijerink and M. Spaans},
   doi = {10.1051/0004-6361/201526538},
   issn = {14320746},
   journal = {A\&A},
   month = {2},
   pages = {A103},
   publisher = {EDP Sciences},
   title = {Consistent dust and gas models for protoplanetary disks: I. Disk shape, dust settling, opacities, and PAHs},
   volume = {586},
   url = {https://ui.adsabs.harvard.edu/abs/2016A&A...586A.103W/abstract},
   year = {2016}
}

@article{Ogihara2018,
   author = {Masahiro Ogihara and Eiichiro Kokubo and Takeru K. Suzuki and Alessandro Morbidelli},
   doi = {10.1051/0004-6361/201832654},
   issn = {14320746},
   journal = {A\&A},
   month = {5},
   pages = {L5},
   publisher = {EDP Sciences},
   title = {Formation of the terrestrial planets in the solar system around 1 au via radial concentration of planetesimals},
   volume = {612},
   url = {https://ui.adsabs.harvard.edu/abs/2018A&A...612L...5O/abstract},
   year = {2018}
}

@article{Lovelace1999,
   author = {R. V. E. Lovelace and H. Li and S. A. Colgate and A. F. Nelson},
   doi = {10.1086/306900},
   issn = {0004-637X},
   issue = {2},
   journal = {ApJ},
   month = {3},
   pages = {805-810},
   publisher = {American Astronomical Society},
   title = {Rossby Wave Instability of Keplerian Accretion Disks},
   volume = {513},
   url = {https://ui.adsabs.harvard.edu/abs/1999ApJ...513..805L/abstract},
   year = {1999}
}

@article{Lovelace2014,
   author = {R. V.E. Lovelace and M. M. Romanova},
   doi = {10.1088/0169-5983/46/4/041401},
   issn = {01695983},
   issue = {4},
   journal = {FlDyR},
   month = {8},
   pages = {041401},
   publisher = {Institute of Physics Publishing},
   title = {Rossby wave instability in astrophysical discs},
   volume = {46},
   url = {https://ui.adsabs.harvard.edu/abs/2014FlDyR..46d1401L/abstract},
   year = {2014}
}

@article{Woitke2024,
   author = {P. Woitke and J. Drazkowska and H. Lammer and K. Kadam and P. Marigo},
   doi = {10.1051/0004-6361/202450289},
   issn = {14320746},
   journal = {A\&A},
   month = {7},
   pages = {A65},
   publisher = {EDP Sciences},
   title = {CAI formation in the early Solar System},
   volume = {687},
   url = {https://ui.adsabs.harvard.edu/abs/2024A%26A...687A..65W/abstract},
   year = {2024}
}

@article{Cecil2024b,
   author = {Michael Cecil and Mario Flock},
   doi = {10.1051/0004-6361/202451175},
   journal = {A\&A},
   month = {11},
   pages = {A171},
   publisher = {EDP Sciences},
   title = {Variability of the inner dead zone edge in 2D radiation hydrodynamic simulations},
   volume = {692},
   url = {http://arxiv.org/abs/2411.05444 http://dx.doi.org/10.1051/0004-6361/202451175},
   year = {2024}
}

@article{Lee2022,
   author = {Eve J. Lee and J. R. Fuentes and Philip F. Hopkins},
   doi = {10.3847/1538-4357/ac8cfe},
   issn = {0004-637X},
   issue = {2},
   journal = {ApJ},
   month = {10},
   pages = {95},
   publisher = {American Astronomical Society},
   title = {Establishing Dust Rings and Forming Planets within Them},
   volume = {937},
   url = {https://ui.adsabs.harvard.edu/abs/2022ApJ...937...95L/abstract},
   year = {2022}
}

@article{Vorobyov2006,
   author = {E. I. Vorobyov and Shantanu Basu},
   doi = {10.1086/507320},
   issue = {2},
   journal = {ApJ},
   month = {7},
   pages = {956-969},
   publisher = {American Astronomical Society},
   title = {The Burst Mode of Protostellar Accretion},
   volume = {650},
   url = {http://arxiv.org/abs/astro-ph/0607118 http://dx.doi.org/10.1086/507320},
   year = {2006}
}

@article{Dzyurkevich2013,
   author = {Natalia Dzyurkevich and Neal J. Turner and Thomas Henning and Wilhelm Kley},
   doi = {10.1088/0004-637X/765/2/114},
   issn = {15384357},
   issue = {2},
   journal = {ApJ},
   month = {3},
   pages = {114},
   publisher = {Institute of Physics Publishing},
   title = {Magnetized accretion and dead zones in protostellar disks},
   volume = {765},
   url = {https://ui.adsabs.harvard.edu/abs/2013ApJ...765..114D/abstract},
   year = {2013}
}

@article{Dzyurkevich2010,
   author = {N. Dzyurkevich and M. Flock and N. J. Turner and H. Klahr and Th Henning},
   doi = {10.1051/0004-6361/200912834},
   issn = {14320746},
   issue = {8},
   journal = {A\&A},
   month = {6},
   pages = {A70},
   publisher = {EDP Sciences},
   title = {Trapping solids at the inner edge of the dead zone: 3-D global MHD simulations},
   volume = {515},
   url = {https://ui.adsabs.harvard.edu/abs/2010A&A...515A..70D/abstract},
   year = {2010}
}

@article{Johns2007,
   author = {Christopher M. Johns‐Krull},
   doi = {10.1086/519017},
   issn = {0004-637X},
   issue = {2},
   journal = {ApJ},
   month = {8},
   pages = {975-985},
   publisher = {American Astronomical Society},
   title = {The Magnetic Fields of Classical T Tauri Stars},
   volume = {664},
   url = {https://ui.adsabs.harvard.edu/abs/2007ApJ...664..975J/abstract},
   year = {2007}
}

@article{Hartmann2016,
   author = {Lee Hartmann and Gregory Herczeg and Nuria Calvet and Lee Hartmann and Gregory Herczeg and Nuria Calvet},
   doi = {10.1146/ANNUREV-ASTRO-081915-023347},
   issn = {0066-4146},
   journal = {ARA\&A},
   month = {9},
   pages = {135-180},
   publisher = {Annual Reviews Inc.},
   title = {Accretion onto Pre-Main-Sequence Stars},
   volume = {54},
   url = {https://ui.adsabs.harvard.edu/abs/2016ARA&A..54..135H/abstract},
   year = {2016}
}

@article{Balbus1991,
   author = {Steven A. Balbus and John F. Hawley},
   doi = {10.1086/170270},
   issn = {0004-637X},
   journal = {ApJ},
   month = {7},
   pages = {214},
   publisher = {American Astronomical Society},
   title = {A Powerful Local Shear Instability in Weakly Magnetized Disks. I. Linear Analysis},
   volume = {376},
   url = {https://ui.adsabs.harvard.edu/abs/1991ApJ...376..214B/abstract},
   year = {1991}
}

@article{Chambers2024,
   author = {John Chambers},
   doi = {10.3847/1538-4357/ad3731},
   issn = {15384357},
   issue = {1},
   journal = {ApJ},
   month = {3},
   pages = {40},
   title = {Large Fluctuations within 1 AU in Protoplanetary Disks},
   volume = {966},
   url = {http://arxiv.org/abs/2403.17126},
   year = {2024}
}

@article{Bai2013,
   author = {Xue Ning Bai and James M. Stone},
   doi = {10.1088/0004-637X/769/1/76},
   issn = {15384357},
   issue = {1},
   journal = {ApJ},
   month = {5},
   pages = {76},
   publisher = {Institute of Physics Publishing},
   title = {Wind-driven accretion in protoplanetary disks. I. Suppression of the magnetorotational instability and launching of the magnetocentrifugal wind},
   volume = {769},
   url = {https://ui.adsabs.harvard.edu/abs/2013ApJ...769...76B/abstract},
   year = {2013}
}

@article{Bae2013,
   author = {Jaehan Bae and Lee Hartmann and Zhaohuan Zhu and Charles Gammie},
   doi = {10.1088/0004-637X/764/2/141},
   issn = {15384357},
   issue = {2},
   journal = {ApJ},
   month = {2},
   pages = {141},
   publisher = {Institute of Physics Publishing},
   title = {Variable accretion outbursts in protostellar evolution},
   volume = {764},
   url = {https://ui.adsabs.harvard.edu/abs/2013ApJ...764..141B/abstract},
   year = {2013}
}

@article{Zhu2009b,
   author = {Zhaohuan Zhu and Lee Hartmann and Charles Gammie},
   doi = {10.1088/0004-637X/694/2/1045},
   issue = {2},
   journal = {ApJ},
   month = {11},
   pages = {1045-1055},
   publisher = {Institute of Physics Publishing},
   title = {Non-steady Accretion in Protostars},
   volume = {694},
   url = {http://arxiv.org/abs/0811.1762 http://dx.doi.org/10.1088/0004-637X/694/2/1045},
   year = {2009}
}

@article{Zhu2010a,
   author = {Zhaohuan Zhu and Lee Hartmann and Charles F. Gammie and Laura G. Book and Jacob B. Simon and Eric Engelhard},
   doi = {10.1088/0004-637X/713/2/1134},
   issn = {15384357},
   issue = {2},
   journal = {ApJ},
   pages = {1134-1142},
   publisher = {Institute of Physics Publishing},
   title = {Long-term evolution of protostellar and protoplanetary disks. I. Outbursts},
   volume = {713},
   url = {https://ui.adsabs.harvard.edu/abs/2010ApJ...713.1134Z/abstract},
   year = {2010}
}

@article{Mohanty2018,
   author = {Subhanjoy Mohanty and Marija R. Jankovic and Jonathan C. Tan and James E. Owen},
   doi = {10.3847/1538-4357/aabcd0},
   issn = {0004-637X},
   issue = {2},
   journal = {ApJ},
   month = {7},
   pages = {144},
   publisher = {American Astronomical Society},
   title = {Inside-out Planet Formation. V. Structure of the Inner Disk as Implied by the MRI},
   volume = {861},
   url = {https://ui.adsabs.harvard.edu/abs/2018ApJ...861..144M/abstract},
   year = {2018}
}

@article{Zhu2010b,
   author = {Zhaohuan Zhu and Lee Hartmann and Charles Gammie},
   doi = {10.1088/0004-637X/713/2/1143},
   issn = {15384357},
   issue = {2},
   journal = {ApJ},
   pages = {1143-1158},
   publisher = {Institute of Physics Publishing},
   title = {Long-term evolution of protostellar and protoplanetary disks. II. Layered accretion with infall},
   volume = {713},
   url = {https://ui.adsabs.harvard.edu/abs/2010ApJ...713.1143Z/abstract},
   year = {2010}
}

@article{Zhu2009,
   author = {Zhaohuan Zhu and Lee Hartmann and Charles Gammie and Jonathan C. McKinney},
   doi = {10.1088/0004-637X/701/1/620},
   issue = {1},
   journal = {ApJ},
   month = {6},
   pages = {620-634},
   publisher = {Institute of Physics Publishing},
   title = {2-D simulations of FU Orionis disk outbursts},
   volume = {701},
   url = {http://arxiv.org/abs/0906.1595 http://dx.doi.org/10.1088/0004-637X/701/1/620},
   year = {2009}
}

@article{Jankovic2022,
   author = {Marija R. Jankovic and Subhanjoy Mohanty and James E. Owen and Jonathan C. Tan},
   doi = {10.1093/mnras/stab3370},
   issn = {13652966},
   issue = {4},
   journal = {MNRAS},
   month = {2},
   pages = {5974-5991},
   publisher = {Oxford University Press},
   title = {MRI-active inner regions of protoplanetary discs – II. Dependence on dust, disc, and stellar parameters},
   volume = {509},
   url = {https://ui.adsabs.harvard.edu/abs/2022MNRAS.509.5974J/abstract},
   year = {2022}
}

@article{Nayakshin2024,
   author = {Sergei Nayakshin and Fernando Cruz Sáenz de Miera and {\'A}gnes Kóspál and Aleksandra Ćalović and Jochen Eislöffel and Douglas N. C. Lin},
   doi = {10.1093/MNRAS/STAE877},
   issn = {0035-8711},
   issue = {2},
   journal = {MNRAS},
   month = {3},
   pages = {1749-1765},
   publisher = {Oxford University Press (OUP)},
   title = {Episodic eruptions of young accreting stars: the key role of disc thermal instability due to Hydrogen ionization},
   volume = {530},
   url = {https://ui.adsabs.harvard.edu/abs/2024MNRAS.530.1749N/abstract},
   year = {2024}
}

@article{Lesur2023,
   author = {G. Lesur and M. Flock and B. Ercolano and M. Lin and C. Yang and J. A. Barranco and P. Benitez-Llambay and J. Goodman and A. Johansen and H. Klahr and G. Laibe and W. Lyra and P. S. Marcus and R. P. Nelson and J. Squire and J. B. Simon and N. J. Turner and O. M. Umurhan and A. N. Youdin and G. Lesur and M. Flock and B. Ercolano and M. Lin and C. Yang and J. A. Barranco and P. Benitez-Llambay and J. Goodman and A. Johansen and H. Klahr and G. Laibe and W. Lyra and P. S. Marcus and R. P. Nelson and J. Squire and J. B. Simon and N. J. Turner and O. M. Umurhan and A. N. Youdin},
   issn = {1050-3390},
   journal = {ASPC},
   pages = {465},
   title = {Hydro-, Magnetohydro-, and Dust-Gas Dynamics of Protoplanetary Disks},
   volume = {534},
   url = {https://ui.adsabs.harvard.edu/abs/2023ASPC..534..465L/abstract},
   year = {2023}
}

@article{Lin1985,
   author = {D. N. C. Lin and J. Papaloizou and J. Faulkner and D. N. C. Lin and J. Papaloizou and J. Faulkner},
   doi = {10.1093/MNRAS/212.1.105},
   issn = {0035-8711},
   issue = {1},
   journal = {MNRAS},
   month = {1},
   pages = {105-149},
   publisher = {Oxford University Press (OUP)},
   title = {On the evolution of accretion disc flow in cataclysmic variables - III. Outburst properties of constant and uniform -alpha model discs.},
   volume = {212},
   url = {https://ui.adsabs.harvard.edu/abs/1985MNRAS.212..105L/abstract},
   year = {1985}
}

@article{Audard2014,
   author = {M. Audard and P. Ábrahám and M. M. Dunham and J. D. Green and N. Grosso and K. Hamaguchi and J. H. Kastner and Á. Kóspál and G. Lodato and M. M. Romanova and S. L. Skinner and E. I. Vorobyov and Z. Zhu},
   doi = {10.2458/azu_uapress_9780816531240-ch017},
   journal = {Protostars and Planets VI},
   pages = {387-410},
   publisher = {University of Arizona Press},
   title = {Episodic Accretion in Young Stars},
   url = {https://ui.adsabs.harvard.edu/abs/2014prpl.conf..387A/abstract},
   year = {2014}
}

@article{Vorobyov2022,
   author = {Eduard I. Vorobyov and Aleksandr M. Skliarevskii and Tamara Molyarova and Vitaly Akimkin and Yaroslav Pavlyuchenkov and {\'A}gnes Kóspál and Hauyu Baobab Liu and Michihiro Takami and Anastasiia Topchieva},
   doi = {10.1051/0004-6361/202141932},
   issn = {14320746},
   journal = {A\&A},
   month = {2},
   pages = {A191},
   publisher = {EDP Sciences},
   title = {Evolution of dust in protoplanetary disks of eruptive stars},
   volume = {658},
   url = {https://ui.adsabs.harvard.edu/abs/2022A&A...658A.191V/abstract},
   year = {2022}
}

@article{Kadam2022,
   author = {Kundan Kadam and Eduard Vorobyov and Shantanu Basu},
   doi = {10.1093/mnras/stac2455},
   issn = {13652966},
   issue = {3},
   journal = {MNRAS},
   month = {11},
   pages = {4448-4468},
   publisher = {Oxford University Press},
   title = {Primordial dusty rings and episodic outbursts in protoplanetary discs},
   volume = {516},
   url = {https://ui.adsabs.harvard.edu/abs/2022MNRAS.516.4448K/abstract},
   year = {2022}
}

@article{Bell1993,
   author = {K. R. Bell and D. N. C. Lin},
   doi = {10.1086/174206},
   journal = {ApJ},
   month = {12},
   pages = {987},
   publisher = {American Astronomical Society},
   title = {Using FU Orionis Outbursts to Constrain Self-Regulated Protostellar Disk Models},
   volume = {427},
   url = {http://arxiv.org/abs/astro-ph/9312015 http://dx.doi.org/10.1086/174206},
   year = {1994}
}

@article{Desch2015,
   author = {Steven J. Desch and Neal J. Turner},
   doi = {10.1088/0004-637X/811/2/156},
   issn = {0004-637X},
   issue = {2},
   journal = {ApJ},
   month = {10},
   pages = {156},
   publisher = {IOP Publishing},
   title = {HIGH-TEMPERATURE IONIZATION IN PROTOPLANETARY DISKS},
   volume = {811},
   url = {https://iopscience.iop.org/article/10.1088/0004-637X/811/2/156 https://iopscience.iop.org/article/10.1088/0004-637X/811/2/156/meta},
   year = {2015}
}

@article{Mignone2007,
   author = {A. Mignone and G. Bodo and S. Massaglia and T. Matsakos and O. Tesileanu and C. Zanni and A. Ferrari},
   doi = {10.1086/513316},
   issn = {0067-0049},
   issue = {1},
   journal = {ApJS},
   month = {5},
   pages = {228-242},
   publisher = {American Astronomical Society},
   title = {PLUTO: A Numerical Code for Computational Astrophysics},
   volume = {170},
   url = {https://ui.adsabs.harvard.edu/abs/2007ApJS..170..228M/abstract},
   year = {2007}
}

@article{Flock2017a,
   author = {M. Flock and S. Fromang and N. J. Turner and M. Benisty},
   doi = {10.3847/1538-4357/835/2/230},
   issn = {0004-637X},
   issue = {2},
   journal = {ApJ},
   month = {2},
   pages = {230},
   publisher = {American Astronomical Society},
   title = {3D Radiation Nonideal Magnetohydrodynamical Simulations of the Inner Rim in Protoplanetary Disks},
   volume = {835},
   url = {https://ui.adsabs.harvard.edu/abs/2017ApJ...835..230F/abstract},
   year = {2017}
}

@article{Malygin2014,
   author = {M. G. Malygin and R. Kuiper and H. Klahr and C. P. Dullemond and Th Henning},
   doi = {10.1051/0004-6361/201423768},
   issn = {14320746},
   journal = {A\&A},
   pages = {A91},
   publisher = {EDP Sciences},
   title = {Mean gas opacity for circumstellar environments and equilibrium temperature degeneracy},
   volume = {568},
   url = {https://ui.adsabs.harvard.edu/abs/2014A&A...568A..91M/abstract},
   year = {2014}
}

@article{Jankovic2021,
   author = {Marija R. Jankovic and James E. Owen and Subhanjoy Mohanty and Jonathan C. Tan},
   doi = {10.1093/mnras/stab920},
   issn = {13652966},
   issue = {1},
   journal = {MNRAS},
   month = {6},
   pages = {280-299},
   publisher = {Oxford University Press},
   title = {MRI-active inner regions of protoplanetary discs. I. A detailed model of disc structure},
   volume = {504},
   url = {https://ui.adsabs.harvard.edu/abs/2021MNRAS.504..280J/abstract},
   year = {2021}
}

@article{Isella2005,
   author = {Andrea Isella and Antonella Natta},
   doi = {10.1051/0004-6361:20052773},
   issue = {3},
   journal = {A\&A},
   month = {3},
   pages = {899-907},
   title = {The Shape of the Inner Rim in Proto-Planetary Disks},
   volume = {438},
   url = {http://arxiv.org/abs/astro-ph/0503635 http://dx.doi.org/10.1051/0004-6361:20052773},
   year = {2005}
}

@article{Shakura1973,
   author = {N. I. Shakura and R. A. Sunyaev},
   issn = {0004-6361},
   journal = {A\&A},
   pages = {337-355},
   title = {Black holes in binary systems. Observational appearance.},
   volume = {24},
   url = {https://ui.adsabs.harvard.edu/abs/1973A&A....24..337S/abstract},
   year = {1973}
}

@article{Levermore1981,
   author = {C. D. Levermore and G. C. Pomraning},
   doi = {10.1086/159157},
   issn = {0004-637X},
   journal = {ApJ},
   month = {8},
   pages = {321-334},
   publisher = {American Astronomical Society},
   title = {A flux-limited diffusion theory},
   volume = {248},
   url = {https://ui.adsabs.harvard.edu/abs/1981ApJ...248..321L/abstract},
   year = {1981}
}

@article{Flock2013,
   author = {M. Flock and S. Fromang and M. González and B. Commerçon},
   doi = {10.1051/0004-6361/201322451},
   issn = {00046361},
   journal = {A\&A},
   month = {12},
   pages = {A43},
   title = {Radiation magnetohydrodynamics in global simulations of protoplanetary discs},
   volume = {560},
   url = {https://ui.adsabs.harvard.edu/abs/2013A&A...560A..43F/abstract},
   year = {2013}
}

@article{Flock2016,
   author = {M. Flock and S. Fromang and N. J. Turner and M. Benisty},
   doi = {10.3847/0004-637x/827/2/144},
   issn = {15384357},
   issue = {2},
   journal = {ApJ},
   month = {8},
   pages = {144},
   publisher = {American Astronomical Society},
   title = {RADIATION HYDRODYNAMICS MODELS OF THE INNER RIM IN PROTOPLANETARY DISKS},
   volume = {827},
   url = {https://ui.adsabs.harvard.edu/abs/2016ApJ...827..144F/abstract},
   year = {2016}
}

@article{Macfarlane2019,
   author = {Benjamin Macfarlane and Dimitris Stamatellos and Doug Johnstone and Gregory Herczeg and Giseon Baek and Huei Ru Vivien Chen and Sung Ju Kang and Jeong Eun Lee},
   doi = {10.1093/mnras/stz1512},
   issn = {13652966},
   issue = {4},
   journal = {MNRAS},
   month = {6},
   pages = {5106-5117},
   publisher = {Oxford University Press},
   title = {Observational signatures of outbursting protostars - I: From hydrodynamic simulations to observations},
   volume = {487},
   url = {https://ui.adsabs.harvard.edu/abs/2019MNRAS.487.5106M/abstract},
   year = {2019}
}

@article{Cleaver2023,
   author = {Jacob Cleaver and Lee Hartmann and Jaehan Bae},
   doi = {10.1093/mnras/stad1784},
   issn = {13652966},
   issue = {4},
   journal = {MNRAS},
   month = {8},
   pages = {5522-5534},
   publisher = {Oxford University Press},
   title = {Magnetically activated accretion outbursts of pre-main-sequence discs},
   volume = {523},
   url = {https://ui.adsabs.harvard.edu/abs/2023MNRAS.523.5522C/abstract},
   year = {2023}
}

@article{Kadam2019,
   author = {Kundan Kadam and Eduard Vorobyov and Zsolt Regály and {\'A}gnes Kóspál and Péter Ábrahám},
   doi = {10.3847/1538-4357/ab378a},
   issn = {15384357},
   issue = {2},
   journal = {ApJ},
   month = {9},
   pages = {96},
   publisher = {American Astronomical Society},
   title = {Dynamical Gaseous Rings in Global Simulations of Protoplanetary Disk Formation},
   volume = {882},
   url = {https://ui.adsabs.harvard.edu/abs/2019ApJ...882...96K/abstract},
   year = {2019}
}

@article{Chrenko2022,
   author = {O. Chrenko and R. O. Chametla and D. Nesvorný and M. Flock},
   doi = {10.1051/0004-6361/202244461},
   issn = {14320746},
   journal = {A\&A},
   month = {10},
   pages = {A63},
   publisher = {EDP Sciences},
   title = {Trapping (sub-)Neptunes similar to TOI-216b at the inner disk rim: Implications for the disk viscosity and the Neptunian desert},
   volume = {666},
   url = {https://ui.adsabs.harvard.edu/abs/2022A&A...666A..63C/abstract},
   year = {2022}
}

@article{Broz2021,
   author = {M. Brož and O. Chrenko and D. Nesvorný and N. Dauphas},
   doi = {10.1038/s41550-021-01383-3},
   issn = {23973366},
   issue = {9},
   journal = {Nat. Astron.},
   month = {9},
   pages = {898-902},
   publisher = {Nature Research},
   title = {Early terrestrial planet formation by torque-driven convergent migration of planetary embryos},
   volume = {5},
   url = {https://ui.adsabs.harvard.edu/abs/2021NatAs...5..898B/abstract},
   year = {2021}
}

@article{Li2022,
   author = {Rixin Li and Yi Xian Chen and Douglas N.C. Lin},
   doi = {10.1093/mnras/stab3677},
   issn = {13652966},
   issue = {4},
   journal = {MNRAS},
   month = {3},
   pages = {5246-5265},
   publisher = {Oxford University Press},
   title = {Dust accumulation near the magnetospheric truncation of protoplanetary discs around T Tauri stars},
   volume = {510},
   url = {https://ui.adsabs.harvard.edu/abs/2022MNRAS.510.5246L/abstract},
   year = {2022}
}

@article{Flock2019,
   author = {Mario Flock and Neal J. Turner and Gijs D. Mulders and Yasuhiro Hasegawa and Richard P. Nelson and Bertram Bitsch},
   doi = {10.1051/0004-6361/201935806},
   issn = {14320746},
   journal = {A\&A},
   month = {10},
   pages = {A147},
   publisher = {EDP Sciences},
   title = {Planet formation and migration near the silicate sublimation front in protoplanetary disks},
   volume = {630},
   url = {https://ui.adsabs.harvard.edu/abs/2019A&A...630A.147F/abstract},
   year = {2019}
}

@article{Gammie1996,
   author = {Charles F. Gammie and Gammie and Charles F.},
   doi = {10.1086/176735},
   issn = {0004-637X},
   journal = {ApJ},
   month = {1},
   pages = {355},
   publisher = {American Astronomical Society},
   title = {Layered Accretion in T Tauri Disks},
   volume = {457},
   url = {https://ui.adsabs.harvard.edu/abs/1996ApJ...457..355G/abstract},
   year = {1996}
}

@article{Steiner2021,
   author = {D. Steiner and L. Gehrig and B. Ratschiner and F. Ragossnig and E. I. Vorobyov and M. Güdel and E. A. Dorfi},
   doi = {10.1051/0004-6361/202140447},
   issn = {14320746},
   journal = {A\&A},
   month = {11},
   publisher = {EDP Sciences},
   title = {Time-dependent, long-term hydrodynamic simulations of the inner protoplanetary disk: I. The importance of stellar magnetic torques},
   volume = {655},
   year = {2021}
}

\begin{appendix} \label{sec:appendix}

\section{Sensitivity test for extrapolated diffusion coefficients} \label{sec:app_extrapolation}
\begin{figure*}[b]
    \centering
         \resizebox{\hsize}{!}{\includegraphics{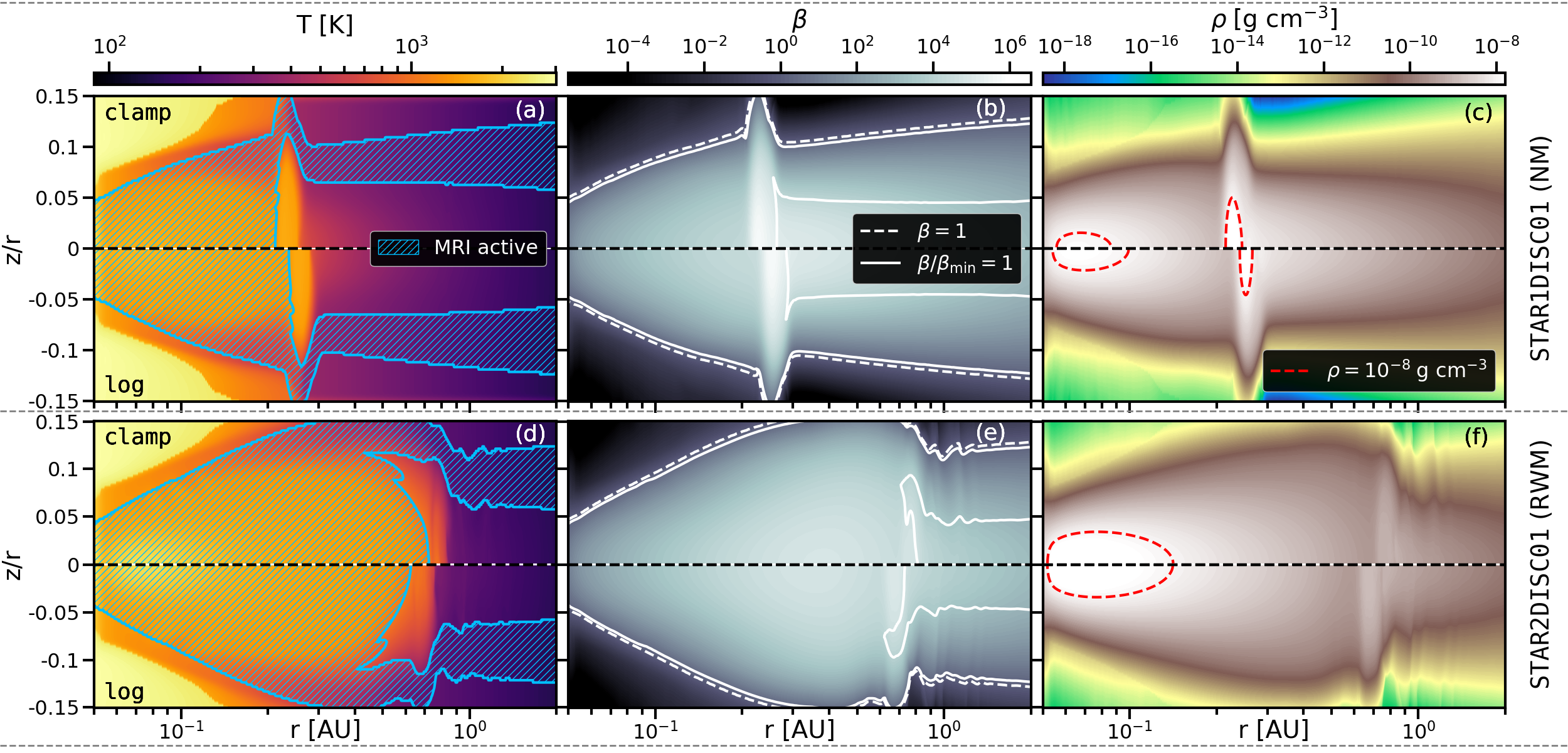}}
    \caption{Comparison of the mid-burst state between models with logarithmic and clamped extrapolation. The upper and lower rows show the temperature, density and $\beta$ maps of the models \texttt{STAR1DISC01} and \texttt{STAR2DISC01}, respectively. The upper and lower hemispheres of each panel display the model with clamped and logarithmic extrapolation, respectively. Analogous to Fig. \ref{fig:structure}, the blue hatched areas in panels a and d mark regions of MRI activity. The red dashed line in panels c and f indicates where the density exceeds the values considered in the lookup table. }
    \label{fig:extrapolation}
\end{figure*}

To test the influence of our chosen extrapolation method of the ambipolar and Ohmic diffusivities from the lookup table provided by \citet{Desch2015}, we conducted two additional simulations of the models confronted in Sect. \ref{sec:res_dichotomy}, \texttt{STAR1DISC01} and \texttt{STAR2DISC01}. For these additional runs, we extrapolated the diffusivities by clamping to the boundary values (constant extrapolation) whenever the input parameters were outside the range covered by the table. We let the models evolve for most of the initial instability cycle to confirm that the burst mode categorisation remains unchanged under the chosen extrapolation method. We compare the mid-burst states (when the MRI active region has reached its largest extent) of the two models, each conducted with both extrapolation methods, in Fig. \ref{fig:extrapolation}. \par
The panels in the first and second columns are analogous to panels a and b of Fig. \ref{fig:structure}. The only quantity that adopts values outside the table's covered range during the simulation is the density, shown in the third column, as it can increase above $10^{-8}~\mathrm{g\;cm^{-3}}$. The red dashed lines in panels c and d indicate the regions where this is the case. \\
The differences arising due to the two different extrapolation methods are minimal. In the NM, the ionisation front can reach slightly larger radii (by less than 0.02 AU) in the logarithmic extrapolation case. The reverse is true in the RWM, although the discernible difference in the extent of the MRI active region in the \texttt{STAR2DISC01} model is also affected by the hydrodynamically unstable motion around the ionisation front. Apart from these small deviations, the evolution of both models is unaffected by the chosen extrapolation method.

\section{Solberg-H{\o}iland criterion} \label{sec:app_solberg}
The Solberg--H{\o}iland criterion used to explore the dichotomy of burst modes in Sect. \ref{sec:res_dichotomy} is expressed as,
\begin{align}
    & C_\mathrm{SH}=N_\mathrm{R}^2+N_\mathrm{Z}^2+\kappa_\mathrm{ef}^2<0 \; , \label{eq:SH1}  %\\
    % & C_\mathrm{SH,2}=\left ( -\frac{\partial P_\mathrm{g}}{\partial Z}\right )\left ( \kappa_\mathrm{ef}^2 \frac{\partial S}{\partial Z} - \frac{1}{R^3}\frac{\partial R^4 \Omega^2}{\partial Z} \frac{\partial S}{\partial R} \right )<0 \;, \label{eq:SH2} 
\end{align}
\noindent where,
\begin{equation}
    \kappa_\mathrm{ef}^2=\frac{1}{R^3}\frac{\partial R^4 \Omega^2}{\partial R} \;,
\end{equation}
\noindent is the squared epicyclic frequency and,
\begin{equation}
    N_\mathrm{R}^2=-\frac{1}{\gamma \rho_\mathrm{g}}\frac{\partial P_\mathrm{g}}{\partial R}\frac{\partial S}{\partial R}, ~~~~N_\mathrm{Z}^2=-\frac{1}{\gamma \rho_\mathrm{g}}\frac{\partial P_\mathrm{g}}{\partial Z}\frac{\partial S}{\partial Z} \;,
\end{equation}
\noindent are the radial and vertical components of the squared Brunt--Väisälä frequency. For the stability analysis, we approximate the entropy with $S=\mathrm{ln(}P_\mathrm{g}\,\rho_\mathrm{g}^{-\gamma})$. $R$ and $Z$ denote the cylindrical radius and height, respectively. Eq. \ref{eq:SH1} allows us to gauge the dynamic stability of the disc, considering both vertical and radial stratification of the specific angular momentum and the entropy. If Eq. \ref{eq:SH1} is satisfied, the respective region can be considered unstable. However, since our models incorporate relatively large values of $\alpha$--prescribed turbulence and the local cooling rate is strongly dependent on radius, density, temperature and opacity, this condition still has to be considered as approximate \citep[e.g.,][]{Latter2016, Teed2021}. For instance, if $C_\mathrm{SH}$ is only marginally smaller than zero (e.g. $0>C_\mathrm{SH}/(\Omega_\mathrm{K}^2)\gtrapprox -10^{-1}$), unstable motion can potentially be suppressed. Formally, the full Solberg--H{\o}iland criterion consists of another condition, which represents a cross-term between radial and vertical buoyancy and rotational shear \citep[][]{Yang2010}. However, due to the complications introduced by our models, we focus in Sect. \ref{sec:res_dichotomy} on the first Solberg--H{\o}iland condition (Eq. \ref{eq:SH1}) as it turns out to be the relevant factor in the investigation of the burst mode dichotomy. \par

\section{Maximum radius of MRI activity given by $\beta$} \label{sec:app_betaradius}
\begin{figure}[t]
    \centering
         \resizebox{\hsize}{!}{\includegraphics{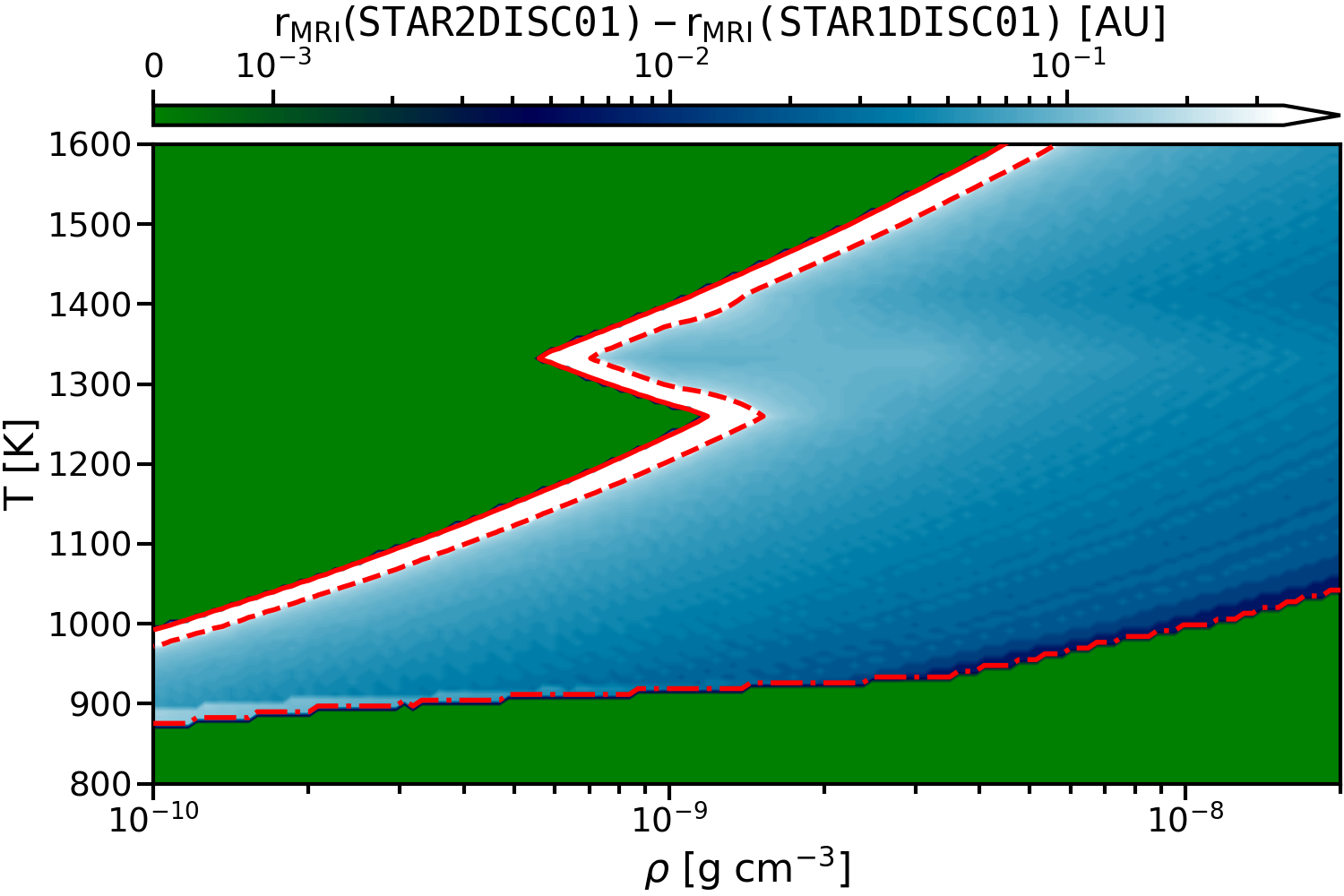}}
    \caption{Difference between the maximum reachable radius of the MRI active region in the midplane between \texttt{STAR2DISC01} and \texttt{STAR1DISC01} in the $\rho - T$ plane. Below the dash-dotted red line, the MRI cannot be activated anywhere in the domain for both models. Above the dashed and solid red lines, the MRI can be made active everywhere for \texttt{STAR2DISC01} and \texttt{STAR1DISC01}, respectively. }
    \label{fig:R_MRI}
\end{figure}
We showed in Sect. \ref{sec:res_dichotomy} that there is an apparent distinct dichotomy between the narrow and wide burst modes. The two analysed models only differ in the prescribed stellar magnetic field strength. As mentioned in Sect. \ref{sec:res_dichotomy}, we do expect the MRI active region to expand further in a slightly more magnetised setting. To analyse how much further the ionisation front can travel in the midplane in the \texttt{STAR2DISC01} model compared to \texttt{STAR1DISC01}, we calculate the maximum radius to which the MRI can be activated, dependent on density and temperature, for both models. We do this by finding the maximum value of $\beta$ for each temperature-density pair, for which the MRI activation conditions are still fulfilled. Since we prescribe the radial profile of the magnetic field strength, we can uniquely identify this $\beta$ value with a radius in our models\footnote{Since we are only interested in the conditions at the midplane, we set the non-thermal ionisation rate to $\zeta=\zeta_\mathrm{RD}=7.6 \cdot 10^{-19}~\mathrm{s^{-1}}$.}. We denote this radius with $r_\mathrm{MRI}$. Fig. \ref{fig:R_MRI} shows the difference in $r_\mathrm{MRI}$ between the two models analysed in Sect. \ref{sec:res_dichotomy}. There are two distinct regions in which the difference becomes zero because the MRI can either be activated nowhere (mostly because the temperatures are too low) or everywhere (because $\beta$ becomes constant far away from the star for the same temperature and density). In between these areas, Fig. \ref{fig:R_MRI} shows that $r_\mathrm{MRI}$ is always larger in \texttt{STAR2DISC01} than in \texttt{STAR1DISC01}, but not by more than $\sim \!\!0.1$~AU. The exception is the thin white area, where the MRI can already be activated everywhere in the more magnetised model, but not yet in the weakly magnetised one. However, in the simulation of \texttt{STAR2DISC01}, the conditions in this area cannot be sustained long enough to advance the ionisation front further ahead of \texttt{STAR1DISC01}. \par
We conclude that although a stronger magnetisation can drive the ionisation front further out on its own, the difference is not large enough to explain the dichotomy observed in the simulations. Therefore, other processes are necessary, which we identify with the Solberg--H{\o}iland-type instability with a contribution from the onset of thermionic emission analysed in Appendix \ref{sec:app_thermionic}.

\section{Effect of dust sublimation} \label{sec:app_thermionic}

\begin{figure}[t]
    \centering
         \resizebox{\hsize}{!}{\includegraphics{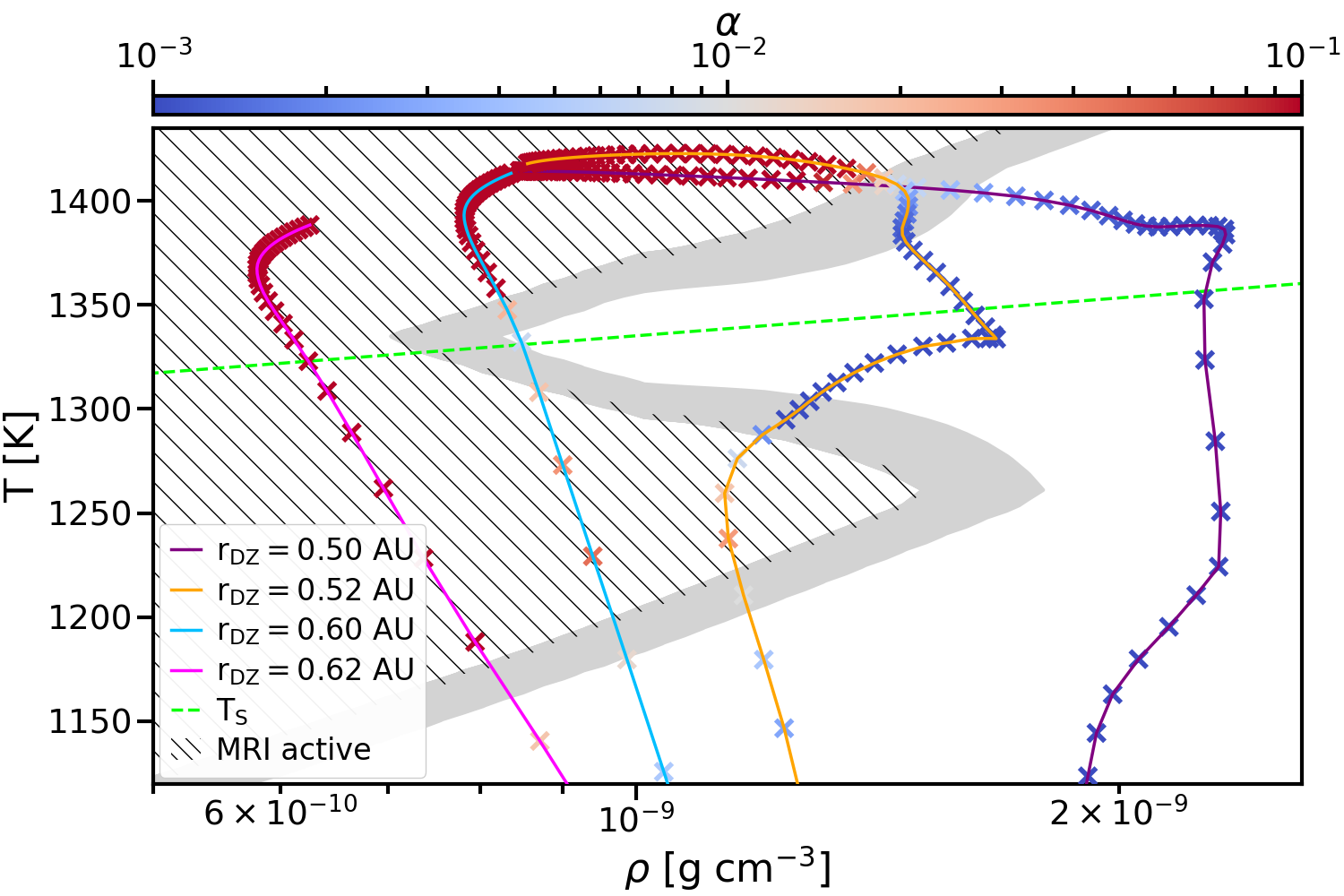}}
    \caption{Same as panel ($\mathrm{d_2}$) of Fig. \ref{fig:SH}, but for four snapshots of \texttt{STAR2DISC01}, equally spaced in time during the period in which the profiles of the midplane density and temperature around the ionisation front are crossing the Z-shaped bend of the MRI activation threshold, shaded in grey. The green dashed line is the sublimation temperature $T_\mathrm{S}$ in our models. The hatched area indicates where the MRI can be sustained.}
    \label{fig:thermionic}
\end{figure}

The study by \citet{Desch2015} clearly showed that thermionic emission can significantly increase the abundance of charge carriers in the inner disc regions by ejecting electrons from dust grain surfaces. It therefore seems reasonable to assume that dust sublimation should leave a clear imprint on the ionisation fraction and, consequently, on the diffusivities and MRI activation criteria. In contrast, \citet{Williams2025} argue that since the ionisation state in the temperature regime where silicates sublimate is dominated by gas-phase reactions, considering sublimation should have no significant influence. However, the lookup tables used in our study do include a feature that can be traced back to the sublimation of silicate grains. It manifests as a Z-shaped curve of the MRI activation transition in the $\rho-T$ plane, visible in panels ($\mathrm{d_1}$) and ($\mathrm{d_2}$) of Fig. \ref{fig:SH} and in Fig. \ref{fig:R_MRI}. \par
When thermionic emission is active, the MRI activation threshold lies at lower temperatures. This transition further aids the outward propagation of the ionisation front at even larger radii than depicted in Fig. \ref{fig:SH}. When the densities and temperatures in the midplane ahead of the ionisation front become small enough to enable thermionic emission contributions, MRI activity is facilitated. Fig. \ref{fig:thermionic} shows panel ($\mathrm{d_2}$) of Fig. \ref{fig:SH} again, but resolving the timeframe in which the midplane conditions around the expanding ionisation front (the radius of which is denoted with $r_\mathrm{DZ}$) cross the Z-shaped feature. \par 
As the front moves outwards, the density around it generally decreases. The orange curve shows that if the density has decreased enough, the MRI can activate beyond the density spike that is pushed ahead of the front. Effectively, this leads to a jump of the front at the midplane towards larger radii (by $\sim\!\!0.08$ AU between the orange and blue lines) as the thermionic emission from dust grains that cannot be sublimated becomes effective. We note that this transition is also occurring above and below the midplane at different radii during the expansion of the MRI active region, which can lead to irregular shapes of the ionisation front, as shown for instance in panel d of Fig. \ref{fig:extrapolation}.

\end{appendix}

\end{document}